# Diffusion bridge with randomized initial and terminal times and its application to fish migration

Hidekazu Yoshioka[1,*] and Mohammed Louriki[2]

[1] Graduate School of Advanced Science and Technology, Japan Advanced Institute of Science and Technology, 1-1 Asahidai, Nomi, Ishikawa, Japan

[2] Mathematics Department, Faculty of Sciences Semalalia, Cadi Ayyad University, Marrakesh, Morocco

* Corresponding author: yoshih@jaist.ac.jp, ORCID: 0000-0002-5293-3246

***Abstract***

We mathematically model the dynamics of the number of migratory fish observed at a fixed location along a river in a random environment. Particularly, as a new approach, we construct a stochastic differential equation that incorporates the influence of environmental factors on the fluctuations in the start and end of migration. The model is a diffusion bridge with a non-Lipschitz diffusion coefficient, called the Cox–Ingersoll–Ross bridge, and has random initial and terminal times arising from time-change, so that the influences of environmental factors can be efficiently incorporated. The well-posedness of the model is first established, which is considered novel and significant in applied mathematics. Second, we estimate the parameters of the model based on the latest multiyear daily data set for the upstream migration of *Plecoglossus altivelis altivelis* (Ayu) by relying on the hypothesis that water temperature affects the migration of the fish, which has been suggested in existing studies. We also explore the application of the proposed model to the challenging task of analyzing environmental DNA data. This study advances the development of a theory of fish migration that is simple yet can take environmental factors into account.

***Statements & declarations***

**Acknowledgments:** We would like to express their gratitude to the Japan Water Agency, Ibi River, and Nagara River General Management Office for providing the fish migration data. We would like to express our sincere gratitude to the members of the Hii Rivet Fisheries Cooperative for their tremendous cooperation during the field survey. The eDNA analysis was outsourced to the Environmental Pollution Research Center Co., Ltd. We thank the two anonymous reviewers for their helpful comments.

**Funding:** This study was supported by the Japan Science and Technology Agency (JST): PRESTO No. JPMJPR24KE.

**Conflict of Interests:** The authors have no relevant financial or nonfinancial interests to disclose.

**Data Availability:** The data will be made available upon reasonable request to the corresponding author.

**Declaration of Generative AI in Scientific Writing:** The author used generative AI to proofread the manuscript.

# 1. Introduction

## 1.1 Background

### 1.1.1 Fish migration

Fish are central to species diversity in aquatic ecosystems in addition to being one of humanity's important nutrition sources [1]. Among fish species, those with a life history involving large-scale migrations, such as salmon and trout, are called migratory fish, and their population dynamics and life histories have been extensively studied [2,3]. Fish migration shapes food chains and nutrient flows in aquatic ecosystems while also directly affecting the abundance or scarcity of fishery resources [4]. Therefore, from multiple perspectives, quantifying the population dynamics of migratory fish is an important research topic

There are several ways to quantify the population dynamics of migratory fish, e.g., sampling surveys [5] and biotelemetry for tracking individual movements [6]; among the many existing methods, one that is possibly noncontact with fish and is widely used in rivers for various fish species is the fixed-point observation of migrants. In earlier studies and surveys, observations have been conducted manually [7] using sonar imaging [8], analyzing images from video cameras [9], and environmental DNA [10]. In either case, some time series data of the number of migrants are obtained, and the main task is its analysis and mathematical modeling.

Previous studies have investigated the influence of environmental factors on the migration of various fish species, which are possibly environment- and region-dependent; high river discharge and salinity have been identified to control the movements of common snook [11]; the catch of chinook salmon increases with proximity to peak river discharge [12]; there is a threshold water temperature above which the juvenile desert fish migrate [13]; the seasonal migration of salmon, trout, and eel around a lake has been found to be triggered by the lake water temperature and its time integration [14,15]; the calendar date has been identified as a driver, dominating local environmental factors, of the upstream migration of several fish species [16]; and an agent-based model based on trajectories of cold-water migratory fish suggested that they migrate while considering the risk of exposure to high water temperatures [17]; watershed-specific socioecological schemes to support life histories of migratory fish have been proposed by focusing on river water temperature [18]. Habitat suitability curves have also been an effective tool for studying fish migration [19].

As described above, there is an accumulation of knowledge about the dynamics of migration in various fish species and environments; however, studies that aim to construct a theory that connects these findings are still lacking. One probable reason for this is that the mathematical methodologies related to fish migration have not yet been fully developed. In particular, there is a lack of established methodologies that can describe the relationship between environmental factors and migration dynamics in mathematical modeling.

### 1.1.2 Diffusion bridge

Movements and migration patterns of biological organisms, including fish, have been effectively modeled by using stochastic differential equations (SDEs), i.e., dynamical systems driven by noise. Particularly,

Brownian bridges [20] have been used for describing animal movements between two spatial points; this method, sometimes called the Brownian bridge movement model, has been applied to the movements of whales in the Atlantic Ocean [21] and the great migration of birds between Africa and Europe [22] and to fish migration, including the estuarine movements of sturgeon [23], juvenile movements of harvested fish species in seascapes [24], and seasonal lake trout management [25].

Diffusion bridges have also been employed in other study areas, such as economics for modeling insider information [26,27], computational fluid dynamics [28,29], biophysics [30], statistical inference [31,32], and diffusion models as a prominent topic in machine learning [33,34]. Thus, diffusion bridges are mathematical models that appear in various research fields. Regarding diffusion bridges, their use in biology is not limited to the movement of organisms between two points in space; Yoshioka [35] applied a diffusion bridge to describing intraday migration of fish, where the initial time and terminal conditions of the bridge are fixed to 0, so that its solution represents the transition of the number of migrating individuals per unit time. The initial and terminal times correspond to start and end times of migration. Typically, as for the abovementioned model, the terminal time of a diffusion bridge is fixed *a priori*; however, this assumption does not necessarily hold true in applications.

Recently, Louriki [36] focused on formulating and mathematically analyzing Brownian bridges and derived models whose terminal times follow a specific probability distribution. However, another issue to consider, from an application standpoint, is the cause of the randomness in the terminal time; for example, this randomness might be a stopping time generated by another stochastic process, which can be an environmental factor in the context of fish migration. At this stage, there are still limited studies on mathematical analysis and application of diffusion bridges with randomness in their terminal (and initial) time. For fish migration, to the best of the authors' knowledge, there are no studies on diffusion bridges driven by random environmental factors.

### 1.2 Aim and contribution

The aim of this paper is to address the two issues mentioned above: modeling the environmental dependence of fish migration and diffusion bridges with random initial and terminal times. The contributions of this study to achieving this aim are described below.

#### 1.2.1 Diffusion bridges with random initial and terminal times

We consider the CIR bridge proposed in Yoshioka [35] as a foundation of the proposed mathematical model. This model and its variants have been applied to modeling unit-time fish count observed at a fixed location along a river and is a diffusion bridge that has a unique nonnegative solution. Its limitation, as partly reviewed above, is that the initial and terminal times, which correspond to the initiation and termination of migration, respectively, are fixed. In contrast, its advantage is the high tractability such that major statistics, such as autocorrelation function and time-dependent cumulants, are available in a closed-form solution due to the nonlinear but affine-SDE nature [35]. Moreover, the model appears from an optimality principle with a biological background [37]. More recently, Yoshioka [38] discussed that CIR bridges serve as a versatile model for describing both intraday and daily fish migration phenomena.

The novelty of the proposed CIR bridge compared to the previous ones is the possible randomness in the migration duration and its timing; the former is a generalization of the latter. More specifically, the proposed model is considered more realistic because the migration durations are not constant in applications. We extend the CIR bridge by introducing an internal, biological clock, which is mathematically represented as a nonnegative and increasing process. This time-change approach leads to a novel CIR bridge whose initial and terminal times are randomized. Well-posedness, i.e., the existence of a unique pathwise solution that is continuous in time and is nonnegative almost surely, of the new CIR bridge is established under a general assumption about the drift and diffusion coefficients, both being possibly singular at the terminal time. The novelty of the proposed CIR bridge compared to the previous ones is the possible randomness in the migration duration; the former is therefore a generalization of the latter.

The qualitative difference between typical diffusion bridges and these CIR bridges is that the former are based on conditioning at terminal time based on Doob's $h$-transform, while the latter use a more heuristic argument. The $h$-transform is a versatile mathematical tool that applies to a wide variety of SDEs [39], even infinite-dimensional [40] and discrete SDEs [41]. In contrast, the heuristic approach is possibly problem specific, and this very point, as mentioned earlier, is what ensures analytical tractability, making it both an advantage and a disadvantage. Bridges arising from an energy minimization principle constitute another stream of studies, particularly in physics [42,43]. The concept of randomized initial and terminal times studied in this paper is considered applicable to a wide range of other diffusion bridges.

#### 1.2.2 Environmental dependence on fish migration

We also apply the proposed CIR bridge to fish migration as a case study. The target fish species is the diadromous fish *Plecoglossus altivelis altivelis* (Ayu), which migrates between seas and rivers and is a major inland fishery resource in a wide area of Japan (e.g., Watanabe et al. [44] and references there). We focus on the upstream migration of juvenile Ayu from the sea to a river that occurs every spring. We construct the biological clock based on the hypothesis that the river water temperature influences the initiation and termination of the upstream migration of Ayu [45,46].

In our application study, we deal with the two data sets about the migration of Ayu. The first data set is the multiyear daily automatic fish count at a weir of a river, and the second data set is the weekly environmental DNA (eDNA) sampling data at an observation point of another river. The eDNA approach has been applied to the quantitative analysis of diverse organisms in environments [47], but the accumulation of knowledge regarding its mathematical modeling is still limited. With the first data set, we conduct computational investigations of this bridge to support sensitivity analysis. The second data are qualitatively different from the first data due to the low temporal resolution and the fact that it does not count migrants. The second application serves as an exploratory case study to advance our investigation into how SDEs, particularly CIR bridges, can be applied to eDNA data. While this application example is still in its early stages, it would provide materials for considering whether and how CIR bridges can be applied to eDNA. Through these contributions, this research will advance both the analysis of new diffusion bridges and the modeling of fish migration.

## 2. CIR bridges

Throughout this paper, we consider a complete filtered probability space $\left(\Omega,\mathcal{F},\left(\mathcal{F}_t\right)_{t\geq 0},\mathbb{P}\right)$ as usual in stochastic calculus for continuous-time processes (e.g., Chapter 5 in Karatzas and Shreve [48]); $\Omega$ is a collection of all events, $\mathcal{F}$ is a collection of all measurable events, $\left(\mathcal{F}_t\right)_{t\geq 0}$ is an increasing family of sigma-algebra, and $\mathbb{P}$ is a probability function.

### 2.1 Existing model without environmental factors

We assume that the total number of migrants is significantly larger than 1 so that we can approximate the population as a real number. We consider an Itô's SDE, which describes the temporal evolution of the unit-time fish count of upstream-migrating fish at a fixed location placed along a river. For a fixed terminal time $T>0$, the CIR bridge proposed in Yoshioka [35] is as follows:

$$\mathrm{d}X_t=\left(a-\frac{r}{T-t}X_t\right)\mathrm{d}t+\sigma\sqrt{X_t}\mathrm{d}B_t,\quad 0<t<T \tag{1}$$

with the initial condition $t=0$. Here, $X=\left(X_t\right)_{0\leq t\leq T}$ is the unit-time fish count (ind/day), $B=\left(B_t\right)_{0\leq t\leq T}$ (day$^{1/2}$) is a one-dimensional standard Brownian motion, $a>0$ (ind/day$^2$) is the source coefficient that represents the supply of migrants from downstream, $\sigma>0$ (ind$^{1/2}$/day) is the volatility that represents fluctuations in the fish count, and $r>0$ (-) is the reversion coefficient that represents the mean-reversion nature of the fish count. The coefficients $a$, $r$, and $\sigma$ are allowed to depend on time $t$. As shown later in **Proposition 1**, the SDE (1) and its generalization investigated in this paper admit a unique pathwise solution that is nonnegative and continuous on $\left[0,T\right]$ and $X_T=0$ almost surely (a.s.). This terminal pinning is due to the unbounded drift coefficient $r/\left(T-t\right)$ when $a$, $r$, and $\sigma$ are constant parameters.

There are at least two drawbacks in the SDE (1). The first drawback is that the time domain $\left[0,T\right]$, which corresponds to the duration of fish migration, is fixed, even though the migration duration varies each year in real data. The second drawback is that it does not consider any environmental factors that affect fish migration. While it would be possible to interpret the influence of environmental factors as being included in the fluctuations of the final term of the SDE (1), this approach makes it difficult to model for specific locations, environments, and fish species.

Previous studies [35,37,38] applied the SDE (1) to daily and intraday fish migration data and demonstrated that the models with finely tuned coefficients can capture the randomness and intermittency observed in the real data of unit-time fish count for both timescales. This SDE has therefore been considered a fundamental model for describing upstream fish migration phenomena, particularly for Ayu.

***Remark 1*** Originally, the diffusion term of the CIR bridge was formulated as $\sigma\sqrt{r/(T-t)X_t}\mathrm{d}B_t$; however, this can be rewritten as in (1) by properly redefining $\sigma$ (ind$^{1/2}$/day$^{1/2}$); therefore, they are essentially the same, but (1) has a visually simpler form. The original formulation is employed in **Section 3** to maintain consistency between the previous and present studies.

## 2.2 Proposed model with environmental factors

### 2.2.1 Formulation

We extend the SDE (1) in the following two ways: first, it is coupled with another stochastic process that represents environmental factors; second, we randomize initial and terminal times based on this new process.

We consider that influences of an environmental factor, such as water temperature (shortened as WT below) or its function, are assumed to be modeled as a nonnegative process $M=(M_t)_{t\geq 0}$ that is defined globally for $t \geq 0$, e.g., $M$ (-) represents the WT or its function. We assume that $M$ is càdlàg, which is not necessarily continuous but right-continuous with left limits. We also define the time integral $\tau=(\tau_t)_{t\geq 0}$ (day) of $M$:

$$\tau_t=\int_0^t M_u \mathrm{d}u,\ \ t\geq 0, \tag{2}$$

which is a continuous and increasing process. We assume that $M$ is independent of the Brownian motion $B$, which implies that environmental factors affect the fish migration but the reverse is not true.

We consider the process $\tau=(\tau_t)_{t\geq 0}$ as a biological clock and assume $\tau_0=0$. Then, for two fixed threshold values $T_1$ (day) and $T_2$ (day) such that $0\leq T_1<T_2<+\infty$, we set the stopping times

$$\theta_i=\inf\left(t>0 \middle| \tau_t>T_i\right),\ \ i=1,2. \tag{3}$$

We have a.s. $\theta_1\leq\theta_2$. Here, $\theta_1$ (day) and $\theta_2$ (day) are the times at which the seasonal fish migration starts and ends, respectively, e.g., threshold values for WT determine the start and end of seasonal migration.

With the processes $M$ and $\tau$, we propose an extended SDE with randomized initial and terminal conditions as follows:

$$\mathrm{d}X_t=\left(a_t-\frac{r_t}{T_2-\tau_t}X_t\right)M_t\mathrm{d}t+\sigma_t\sqrt{M_tX_t}\mathrm{d}B_t,\ \ \theta_1<t<\theta_2 \tag{4}$$

subject to $X_{\theta_1}=0$. The differences between the two SDEs (1) and (4) are that the initial and terminal times are randomized and the environmental factor $M_t$ appears in the latter. At first glance, the appearance of $M_t$ is unnatural. However, its role becomes clearer if we employ the following transformation of variables:

$$X_t=Y_s \ \text{ and } \ s=\tau_t=\int_0^t M_u\mathrm{d}u. \tag{5}$$

Indeed, we formally have $\mathrm{d}s=M_t\mathrm{d}t$, and hence, the SDE (4) is rewritten as follows:

$$\mathrm{d}Y_s = \left( \hat{a}_s - \frac{\hat{r}_s}{T_2 - s} Y_s \right) \mathrm{d}s + \hat{\sigma}_s \sqrt{Y_s} \mathrm{d}\hat{B}_s, \quad T_1 < s < T_2 \tag{6}$$

subject to $Y_{T_1} = 0$, where $s$ (day) represents biological time, and we set $\hat{a}_s = a_t$, $\hat{\sigma}_s = \sigma_t$, and $\hat{r}_s = r_t$, and $\hat{B} = \left(\hat{B}_s\right)_{s \geq 0}$ is some one-dimensional standard Brownian motion, which exists under certain conditions **(Proof of Proposition 1)**. The transformed SDE (6) has the same form as the original one (1). Thus, the nonlinear SDE (4) arises from the original model defined in biological time in conjunction with a proper time change. A mathematical issue that needs to be investigated regarding the proposed model is under which conditions such time changes are permitted.

***Remark 2*** The proposed SDE (4) reduces to the original model (1) if $M$ is identically a positive constant.
***Remark 3*** A seemingly simpler way to model randomized terminal time is to use the following SDE:

$$\mathrm{d}X_t = \left( a - \frac{r}{\zeta - t} X_t \right) \mathrm{d}t + \sigma \sqrt{X_t} \mathrm{d}B_t, \quad 0 < t < \zeta, \tag{7}$$

where $\zeta > 0$ (day) is a random variable. This SDE has a theoretical difficulty in simulating its sample path because it uses the value of $\zeta$, which is random and is therefore *a priori* unknown for observers. However, this model has another potential interpretation, in which $\zeta$ is a genetically coded quantity that may vary among generations. In this case, this $\zeta$ is not directly observable by humans. Although this deviates from the main point of this paper, it might also be possible to interpret this difference as leading to information asymmetry.

### 2.2.2 Well-posedness

We establish sufficient conditions for the existence of a unique pathwise solution to our CIR bridge. To proceed, we set the following assumption.

***Assumption 1*** *The process $M$ is nonnegative and càdlàg. Assume $M_t > 0$ Lebesgue-a.e. a.s. and that M is locally bounded a.s. on $(\theta_1, \theta_2)$. Furthermore, assume $\int_0^\infty M_t \mathrm{d}t = \infty$ a.s..*

Under **Assumption** 1, $\tau_t$ is strictly increasing on $(\theta_1, \theta_2)$; $\theta_1 < \theta_2 < \infty$ a.s. follows from $T_1 < T_2 < \infty$ and $\int_0^\infty M_t \mathrm{d}t = \infty$. We also set the following assumption about the coefficients in the SDE (6).

***Assumption 2***

*(i) At all $s \in [T_1, T_2)$, $\hat{a}_s$ is bounded and continuous, and it behaves near $s = T_2$ as $\hat{a}_s = O\left(\left(T_2 - s\right)^\alpha\right)$ for some $\alpha > -1$.*

*(ii)* *At all* $s \in [T_1, T_2)$ *,* $\hat{\sigma}_s$ *is bounded and continuous, and it behaves near* $s = T_2$ *as* $\hat{\sigma}_s = O\left((T_2 - s)^{\beta}\right)$ *for some* $\beta > -\frac{\alpha}{2} - 1$*, where* $\alpha$ *is specified in (i).*

*(iii)* *At all* $s \in [T_1, T_2]$*,* $\hat{r}_s$ *is continuous.*

This **Assumption 2**, which allows for unbounded coefficients, is weaker than the corresponding ones employed in the previous studies [35,37,38] and is therefore mathematically more general.

Now, we state the main mathematical result of this paper.

***Proposition 1***

*Under **Assumptions 1-2**,* $\theta_1 < \theta_2$ *holds true a.s., and the SDE (4) admits a pathwise unique solution in* $(\theta_1, \theta_2)$ *that is a.s. nonnegative and satisfies* $X_{\theta_1} = X_{\theta_2} = 0$ *.*

***Remark 4*** According to **Proposition 1**, the coefficients $\hat{a}, \hat{\sigma}$ are allowed to blow up at the terminal time $s = T_2$ if the speed is not fast enough. The blow-up of $\hat{\sigma}$ was supposed in Yoshioka [35], while $\hat{a}$ was assumed to be bounded. The application study in **Section 3** implies that this is not the case for Ayu; however, it is uncertain whether the same applies to other fish species.

### 2.2.3 Examples of $M$

We present two examples of the process $M$ as an (aggregated) environmental factor. The first one is the random step process (recall that $M$ is allowed to be discontinuous):

$$M_t = \begin{cases} 0 & (0 \le t < \theta_1) \\ 1 & (\theta_1 \le t < \hat{\theta}_2) \\ \omega & (t \ge \hat{\theta}_2), \end{cases} \tag{8}$$

where $\omega > 1$ (-) is a constant, $\theta_1 = \inf(t > 0 | \tau_t = 0) = \inf(t > 0 | M_t = 0)$, $\theta_2 = \inf(t > \theta_2 | \tau_t = T_2)$, and $\hat{\theta}_2 = \inf(t > \theta_1 | w_t > \bar{w})$ with $w = (w_t)_{t \ge 0}$ (°C) being WT and $\bar{w}, \underline{w} > 0$ (°C) being threshold values. Here, we set $T_2 > T_1 = 0$ . For example, $w$ follows a diffusion process. In this case, migration starts when $w$ enters $(\underline{w}, \bar{w})$ of the preferable WT for migration of the target fish species: $t = \theta_1$ . After $w$ first left $(\underline{w}, \bar{w})$, the migration rapidly proceeds toward the end due to $\omega > 1$ ($t = \hat{\theta}_2$), and it terminates at $t = \theta_2$ . From a biological viewpoint, the process $M$ in (8) is a step function for determining the duration $(\theta_1, \theta_2)$ of migration, during which environmental conditions, such as WT and water level, are suited to the migration of a target fish species. The case $\omega = 1$ is the simplest because the speeds of physical and biological clocks coincide. The limit $\omega \to +\infty$ corresponds to the situation where the fish migration is immediately terminated at time $\hat{\theta}_2$. Note that the model (8) needs further validation in the future because we are not aware of such data for Ayu. Instead, sensitivity of the parameter $\omega$ is computationally studied later.

The second one is given by

$$M_t = \max\{w_t - \underline{w}, 0\}. \tag{9}$$

Then, the process $\tau$ is the accumulated WT or its rectified version (the right-hand side being nondimensionalized with a coefficient "1" having the unit of inverse temperature). In this case, the migration starts at which the WT exceeds the threshold value $\bar{w}$ and ends at another biological threshold. Technically, the right-hand side of (9) can be replaced by a nondecreasing function of $w_t - \bar{w}$. A drawback of using (9) is the fact that the physical meaning of a summation or integration of "(water) temperature", which is an intensive variable, is unclear.

Finally, in view of **Assumption 1** and **Proposition 1**, the process $M$ and $\theta_1, \theta_2$ should be designed as a positive function of environmental factors on $(\theta_1, \theta_2)$, which is satisfied at least in the first example above.

### 2.2.4 Numerical method

We present a numerical method for computing our CIR bridge in a physical time axis. A difficulty in applying a common numerical method, such as the Euler–Maruyama method, to the SDE (4) is that the coefficients blow up at the terminal time and the diffusion coefficient is non-Lipschitz, i.e., $\sqrt{X_t}$, and it experiences an extremely slow convergence in computation (e.g., Hefter and Herzwurm [49]); with these difficulties, one may obtain negative numerical solutions or even computational failure.

Yoshioka [35] demonstrated that these drawbacks can be resolved by using the integrated variance implicit (iVi) scheme proposed in Abi Jaber [50][1]. This numerical method is unconditionally stable, generates nonnegative numerical solutions for any time increment (Theorem 1.3 [50]), and has been found to work for the SDE (1) and related models [51,52].

Here, we present a numerical scheme for the SDEs (4). First, we set a time increment $\Delta t > 0$ of physical time $t$ and time steps $t_k = k\Delta t$ ($k = 1,2,3,...,K$), where $K$ is a large natural number. Second, we set the recursion for $\tau$ at each $t_k$ as follows, where $\hat{\tau}_0 = 0$:

$$\hat{\tau}_{k+1} = \hat{\tau}_k + \hat{M}_k \Delta t, \quad k = 0,1,2,..., \tag{10}$$

where we assume that each $\hat{M}_k$, which is the approximation of $M_{t_k}$, is already available. In our application, we construct $M$ as a function of an Ornstein–Uhlenbeck process, which can be easily simulated by using a classical Euler–Maruyama method. We set

[1] The published version is available at the following link but seems not be accessible to everyone. https://www.risk.net/cutting-edge/7961711/simulation-of-heston-made-simple

$$k_i = \inf\left(k \middle| \hat{\tau}_k > T_i\right), \quad i = 1,2 . \tag{11}$$

Assume that $k_1 < k_2$, which is possible in practice if we choose a sufficiently small $\Delta t$. Empirically, choosing $\Delta t$ to be smaller than 1/1,000 to 1/10,000 of the migration duration seems to be enough. Then, we obtain the approximations $\tilde{\theta}_i = t_{k_i}$ of $\theta_i$ ($i = 1,2$).

Second, we approximate $X$ at the restricted grid points $t_k$ ($k = k_1, k_1+1, ..., k_2$) by discretizing the SDE (4) in $\left(t_{k_1}, t_{k_2}\right)$. We set $t_{k+1/2} = \left(t_k + t_{k+1}\right)/2$. The numerical solution at time $t_k$ is denoted by $\hat{X}_k$, where $\hat{X}_{k_1} = 0$. We set $\tilde{a}_k = a_{t_{k+1/2}} M_{t_k} \geq 0$, $\tilde{R}_k = r_{t_{k+1/2}} M_{t_k} / \left(T_2 - \hat{\tau}_k\right) \geq 0$, $\tilde{\sigma}_k = \hat{\sigma}_{t_{k+1/2}} \sqrt{M_{t_k}} \geq 0$, and the composite parameters

$$\phi_k = \hat{X}_k \frac{1 - \exp\left(-\tilde{R}_k \Delta t\right)}{\tilde{R}_k} + \frac{\tilde{a}_k}{\tilde{R}_k}\left(\Delta t - \frac{1 - \exp\left(-\tilde{R}_k \Delta t\right)}{\tilde{R}_k}\right) \geq 0 \tag{12}$$

and

$$\psi_k = \tilde{\sigma}_k \frac{1 - \exp\left(-\tilde{R}_k \Delta t\right)}{\tilde{R}_k} \geq 0 . \tag{13}$$

We also set random variables

$$\hat{U}_{k,k+1} \sim IG\left(\phi_k, \left(\phi_k \psi_k^{-1}\right)^2\right) \text{ and } \hat{Z}_{k,k+1} = \frac{1}{\psi_k}\left(\hat{U}_{k,k+1} - \phi_k\right). \tag{14}$$

Here, $IG(\mu, \lambda)$ represents the inverse gamma distribution with the probability density function

$$f(z; \mu, \lambda) = \sqrt{\frac{\lambda}{2\pi z^3}} e^{-\frac{\lambda (z-\mu)^2}{2\mu^2 z}}, \quad z > 0 . \tag{15}$$

We assume that $\left\{\hat{U}_{k,k+1}\right\}_{k=0,1,2,...,K-1}$ are independent and identically distributed samples. Finally, we set the recursion:

$$\hat{X}_{k+1} = \hat{X}_k + \tilde{a}_k \Delta t - \tilde{R}_k \hat{U}_k + \tilde{\sigma}_k \hat{Z}_{k,k+1}, \quad k = k_1, k_1+1, ..., k_2 - 1 \tag{16}$$

if $M_{t_k} > 0$ and $\hat{X}_{k+1} = 0$ if $M_{t_k} = 0$. The discretization scheme (16) is an adaptation of **Algorithm 1** in Abi Jaber [50] to our case. The unconditional nonnegativity of the computed $\hat{X}_k$ is due to the specific choices of the right-hand side of (16) as proven in Theorem 1.3 of Abi Jaber [50]; although not presented here, the coefficients and random numbers there come from a Riccati equation associated with the discretized process, which explains why the scheme successfully handles the CIR-type SDEs.

***Remark 5*** Another way to simulate our CIR bridge is to exploit the equivalence between the two SDEs (4) and (6); the SDE (6) is solved first to approximate $Y$, and then $X$ on a random, nonuniform grid. We do not choose this method because of the requirement of the two different grids.

## 3. Applications

### 3.1 Target fish: Ayu

Ayu, our target fish species, has a yearly life cycle that involves migration between the ocean and rivers (details are left to the literature, e.g., Tsukamoto and Uchida [53]). In spring, juveniles migrate from the ocean to rivers, spend the summer in the middle reaches of rivers, spawn in the lower reaches of rivers in autumn, and then die. After hatching, larvae travel downstream to the ocean, where they spend the winter. Due to this life cycle, Ayu has a simple annual life cycle with limited age structure, establishing its position as a model species in the study of migratory fish.

### 3.2 Nagara River study: fish count

#### 3.2.1 Site description and data

The first application site of this study is the Nagara River, a first-class river that flows through the Tokai region of Japan and empties into Ise Bay and then the Pacific Ocean (**Figure 1**). The Nagara River has the Nagara River Estuary Barrage located 5.4 (km) from its mouth (35°4'48"N, 136°41'51"E), and a video recording device is installed in a part of its fish ladder. Using this device, the number of juvenile Ayu migrating upstream was recorded every year from February to June at a 10-minute resolution. Additionally, Oyabu Big Bridge Observation Station (called Oyabu in the sequel: 35°17'48"N, 136°40'16"E), authorized by the Ministry of Land, Infrastructure, Transport and Tourism, is located approximately 31.2 (km) from the river mouth, where hourly WT is measured. We did not address the relationship between fish migration and water level based on site characteristics because of the regular tidal dominance over flood disturbances[2].

In the Nagara River, Ayu is an important inland fishery resource, and traditional fishing methods such as cormorant fishing are practiced there, along with a unique culture based on the fish [54]. This river has also been a study site for biodiversity loss due to anthropogenic pressures [55], and watershed-scale studies such as WT surveys [56] and growth-strategy investigations of Ayu [57] have recently been conducted. Thus, the Nagara River is an ideal site for studying the coexistence of humans and the environment, as well as the population dynamics of Ayu as part of the ecosystem and as a fishery resource. **Figure 2** shows the observed fish count data of Ayu and daily averaged WT (called daily WT in the sequel) (°C) at Oyabu from 2016 to 2025.

[2] See the data at Sotohama observation station: Last accessed on May 278, 2026.
https://www1.river.go.jp/cgi-bin/SiteInfo.exe?ID=305091285515150

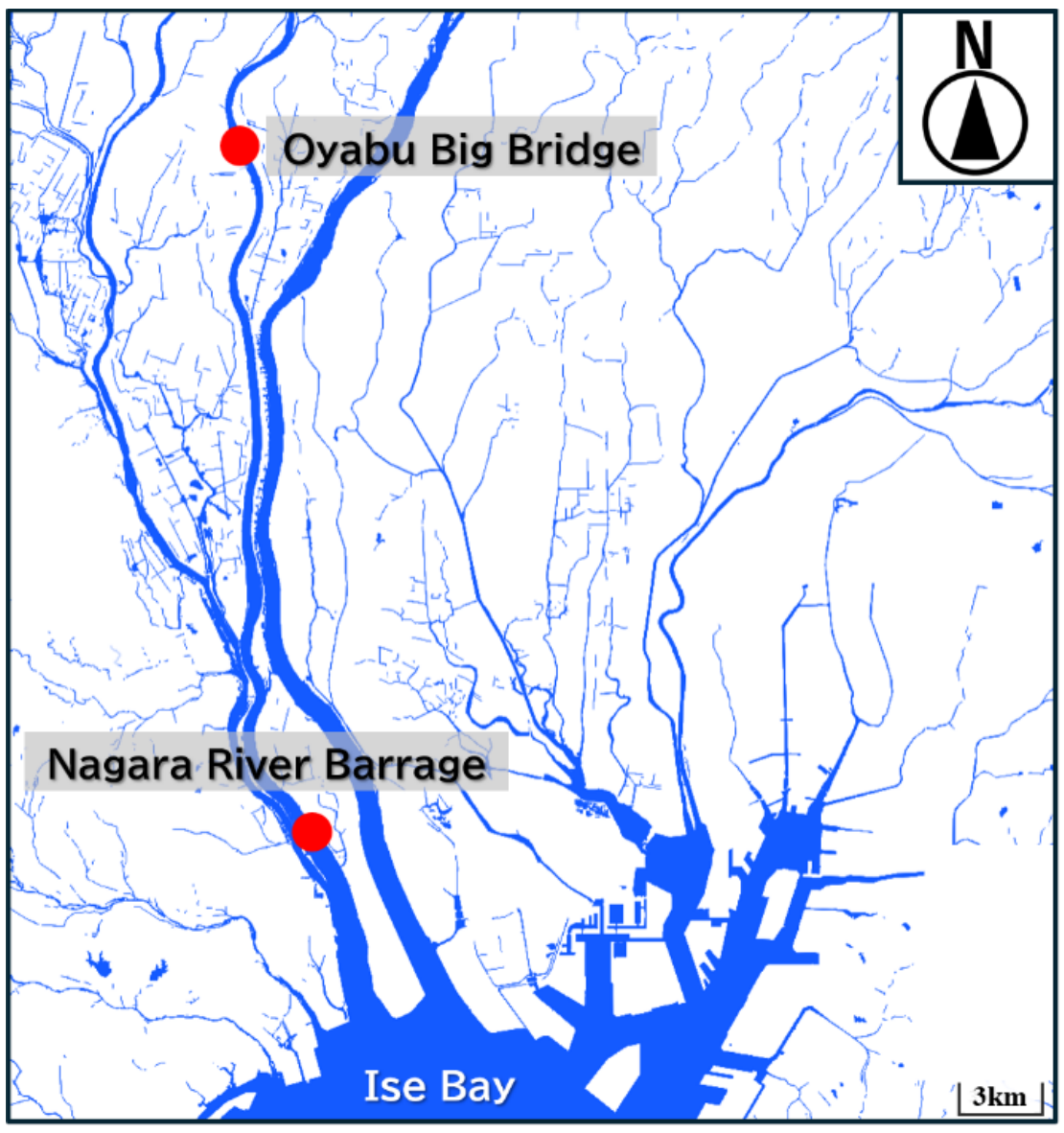

**Figure 1.** Map of the study site in the Nagara River (modified from Chiri-in Chizu Vector: https://github.com/gsi-cyberjapan/gsimaps-vector-experiment).

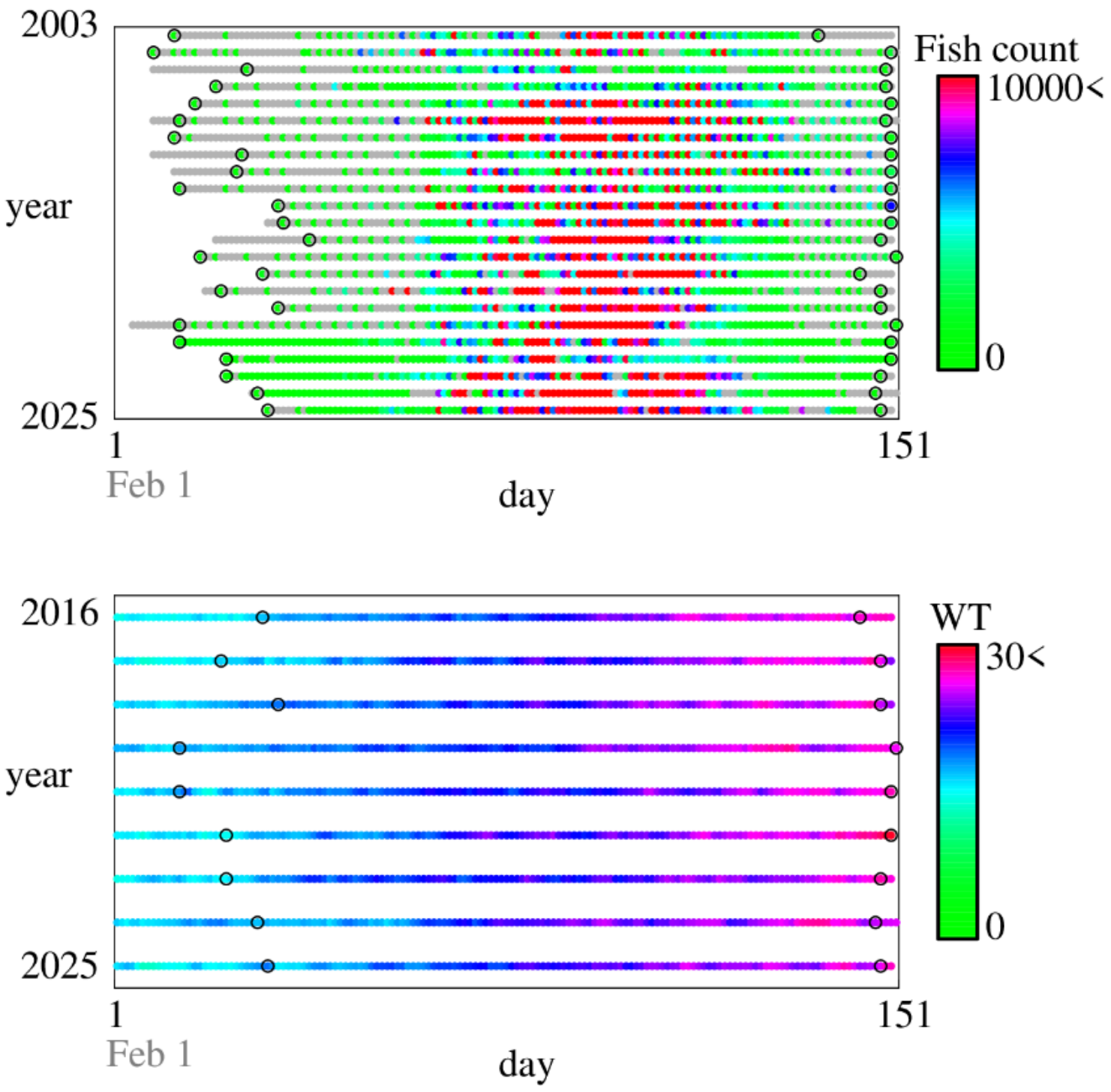

**Figure 2.** Daily data of (a) fish count of Ayu at the Nagara River Estuarine Barrage from 2003 to 2025 and (b) daily WT (°C) at Oyabu from 2016 to 2025. Gray plots represent the zero count. Black circles represent the start and end dates of the migration of Ayu each year.

### 3.2.2 Parameter estimation

We consider the WT at Oyabu to be an environmental factor. This is due to the empirical finding in the literature that the WT significantly influences the upstream migration of Ayu [46,47]. Moreover, as discussed in Paíz et al. [14], the approach based on a single environmental factor would be better transferable to other river environments than those that need multiple factors. In addition, as reviewed in Weedop et al. [16], calendar time has been suggested to be a dominant factor that shapes fish migration, and hence, adding factors indiscriminately may not be promising.

We also face another issue, which is the data availability of WT, where the record of WT at Oyabu is not always sufficient for the modeling purpose, e.g., more than 15% of data points are missing in and before 2015[3]. WT analysis is performed with the data since 2016. Yoshioka [38] identified the CIR bridge normalized with two quantities: migration duration and total number of migrants with some parametric versions of $a$ and $\sigma$ (**Appendix B**). In this model, migration duration in each year was given *a priori*, and their temperature dependence has not been investigated. Considering these data and modeling circumstances, in the first application of this study, we reinterpret the previous model [38] from the viewpoint of random migration durations, which is possible by using the first model (8) of the process $M$ with $\omega = 1$ as explained below.

We explain the above model with a normalization procedure. Assume temporally that the initial and terminal times $\theta_1, \theta_2$ (based on first/last non-zero observations) and the total number of migrants $S = \int_{\theta_1}^{\theta_2} X_u \mathrm{d}u > 0$ (ind) are given, e.g., observed values. For simplicity, we write $T = \theta_2 - \theta_1$. In Yoshioka [38], the SDE (4), by assuming $M \equiv 1$, was normalized by using the following nondimensional variables: $\bar{t} = (t - \theta_1)/T$ and $\bar{X}_{\bar{t}} = X_t T_{\text{day}} / S$ with $T_{\text{day}} = 1$ day, and hence $\tau_t = \bar{\tau}_{\bar{t}} = T\bar{t}$. The normalized SDE reads, in the original form (**Remark 1**), as follows:

$$\frac{S}{T_{\text{day}}} \mathrm{d}\bar{X}_{\bar{t}} = \left( a_t - \frac{r_t}{T - T\bar{t}} \frac{S}{T_{\text{day}}} \bar{X}_{\bar{t}} \right) T \mathrm{d}\bar{t} + \sigma_t \sqrt{\frac{S}{T_{\text{day}}} \frac{r_t}{T - T\bar{t}} \bar{X}_{\bar{t}}} \sqrt{T} \mathrm{d}\bar{B}_{\bar{t}}, \quad 0 < \bar{t} < 1, \tag{17}$$

where $\bar{B}$ is a 1-D standard Brownian motion. Rearranging (17) yields

$$\mathrm{d}\bar{X}_{\bar{t}} = \left( \frac{T_{\text{day}} T}{S} a_t - \frac{r_t}{1 - \bar{t}} \bar{X}_{\bar{t}} \right) \mathrm{d}\bar{t} + \sigma_t \sqrt{\frac{T_{\text{day}}}{S}} \sqrt{\frac{r_t}{1 - \bar{t}} \bar{X}_{\bar{t}}} \mathrm{d}\bar{B}_{\bar{t}}, \quad 0 < \bar{t} < 1. \tag{18}$$

Yoshioka [38] applied this normalization procedure to daily fish count data each year by assuming that the following normalized quantities are common among all the years: $\bar{a}_{\bar{t}} = S^{-1} T_{\text{day}} T a_t$, $\bar{\sigma}_{\bar{t}} = \sigma_t \sqrt{S^{-1} T_{\text{day}}}$, and $r_{\bar{t}} = r_t \,(= \text{constant})$.

To investigate our CIR bridge along the physical timeline but not in nondimensional coordinates, we need to replace $S$ with some available value that serves as a good representative of all the data. We specify this $S$ so that the identified model in the physical timeline has the empirical average of the total number of migrants on average. More specifically, let $T_{\mathrm{emp}} = \mathbb{E}[\theta_2 - \theta_1]$ be the empirical average of migration durations. Then, we set

$$\int_0^{T_{\mathrm{emp}}} \mathbb{E}[X_u]\big|_{T=T_{\mathrm{emp}}} \mathrm{d}u = S_{\mathrm{emp}}, \tag{19}$$

where the initial time is formally shifted to 0. By using the average formula of (18) (**Appendix B**), we can simplify (19) to

$$S\frac{AT_{\mathrm{emp}}}{T_{\mathrm{day}}}\int_0^{T_{\mathrm{emp}}}\left(\frac{t}{T_{\mathrm{emp}}}\right)^m\left(1-\frac{t}{T_{\mathrm{emp}}}\right)^n \mathrm{d}\left(\frac{t}{T_{\mathrm{emp}}}\right) = S_{\mathrm{emp}}, \tag{20}$$

where the parameters $A, m, n > 0$ are available in Yoshioka [38] (see also **Section B.1** in **Appendix**). Then, we can explicitly find $S$ from (20) because the integral on the left-hand side of (20) is a beta function with the degree $(m+1, n+1)$. **Table 1** summarizes the nominal parameter values of the model, showing that we have approximately $S = S_{\mathrm{emp}}$. In the rest of this paper, we assume $S$ to be $S_{\mathrm{emp}}$ for the (de)normalization.

[3] Last accessed on May 25, 2026. http://www1.river.go.jp/cgi-bin/SrchWquaData.exe?ID=405091285502120&KIND=5&PAGE=0

The difference between the maximum and minimum migration durations from 2003 to 2025 is 32 days (maximum 143 (day) and minimum 111 (day)), and their deviations from the average of 127.8 days are 15% and 12%, respectively. Additionally, we have the start and end dates of the upstream migration of Ayu and daily WT on these dates at Oyabu each year, as summarized in **Table 2**. Some of the end dates are the same as those of the end of the observation window (June 30); therefore, the estimation of the end date may have been biased to be earlier. For example, there may be nonzero fish count after June 30. We found a low correlation between migration duration and WT difference ($R^2$ value 0.113), and hence, they are not considered in a linear relationship. In contrast, **Table 2** shows that there is an approximately 14 (°C) average difference between the daily WTs on the start (9.07 (°C) on average) and end dates (23.2 (°C) on average) of migration. The analysis from the restricted period would restrict its accuracy as pointed out in the comment, and therefore the accumulation of more reliable data is desired. **Table B1** shows the linear regression results of daily WT from February to June at Oyabu along with the corresponding $R^2$ values each year, demonstrating that the linear regression serves well ($R^2$ values are approximately 0.9 for all cases).

The two investigation results above, the WT difference between the start and end dates of migration and the linear relationship between the WT and time, suggest that the model (8) with $\underline{w} = 9.07$ (°C) and $\overline{w} = 23.2$ (°C) can serve as a baseline. For the deterministic WT case, we choose the linear model:

$$w_t = \hat{w}_t := \underline{w} + \frac{t - t_{\mathrm{start}}}{T_{\mathrm{emp}}}\left(\overline{w} - \underline{w}\right), \quad t > 0, \tag{21}$$

where $t_{\mathrm{start}} = 20.7$ (day) is the average of start dates from 2003 to 2025. The slope of (21) is defined by $\hat{\kappa} = \left(\overline{w} - \underline{w}\right) / T_{\mathrm{emp}}$. We also investigate its stochastic version where $w$ is an Ornstein–Uhlenbeck process:

$$\mathrm{d}w_t = -a^{(w)}\left(w_t - \hat{w}_t\right)\mathrm{d}t + b^{(w)}\mathrm{d}N_t, \quad t > 0 \tag{22}$$

with $w_0$ determined by substituting $t=0$ into the deterministic model (21), where $N=(N_t)_{t\geq 0}$ is a 1-D standard Brownian motion that is independent of those driving CIR bridges, and $a^{(w)}=0.1884$ (1/day) and $b^{(w)}=0.8533$ (°C/day$^{1/2}$) are parameters identified from the daily WT data (**Appendix B**). The SDE (22) formally reduces to (21) when $a^{(w)}\to+\infty$. The SDE (22) is examined in this paper to investigate how noise in WTs would affect fish migration dynamics. With the above preparations in place, it is now possible to generate sample paths for CIR bridges for $\omega>1$.

**Table 1.** Parameter values for the nominal case (see also **Section B.1** in **Appendix**).

| Parameter | Value | Remark |
|---|---|---|
| $A$ (-) | 1.898E+04 | Taken from Yoshioka [39] |
| $V$ (-) | 1.475E+04 | |
| $m$ (-) | 1.137E+01 | |
| $n$ (-) | 8.361E+00 | |
| $p$ (-) | 1.391E+01 | |
| $q$ (-) | 1.032E+01 | |
| $r$ (-) | 6.190E+01 | |
| $T_{\mathrm{emp}}$ (day) | 1.278E+02 | |
| $S_{\mathrm{emp}}$ (ind) | 8.362E+05 | |
| $S$ (ind) | 8.861E+05 | |

**Table 2.** Start and end dates and fish count of the upstream migration of Ayu and WT on these dates. Migration duration and WT differences each year are also presented.

| Year | Date (Start) | Date (End) | Fish count | Duration (day) | WT (°C) (Start) | WT (°C) (End) | WT (°C) difference |
|---|---|---|---|---|---|---|---|
| 2025 | 2-Mar | 28-Jun | 1,318,281 | 119 | 10.8 | 21.8 | 11.0 |
| 2024 | 28-Feb | 26-Jun | 1,236,102 | 120 | 9.40 | 20.7 | 11.3 |
| 2023 | 22-Feb | 28-Jun | 852,596 | 127 | 7.66 | 25.2 | 17.6 |
| 2022 | 22-Feb | 30-Jun | 224,397 | 129 | 7.40 | 28.9 | 21.5 |
| 2021 | 13-Feb | 30-Jun | 403,459 | 138 | 9.77 | 24.6 | 14.8 |
| 2020 | 13-Feb | 30-Jun | 812,342 | 139 | 8.79 | 22.4 | 13.6 |
| 2019 | 4-Mar | 28-Jun | 592,439 | 117 | 11.5 | 21.0 | 9.56 |
| 2018 | 21-Feb | 28-Jun | 847,565 | 128 | 8.65 | 22.8 | 14.1 |
| 2017 | 1-Mar | 24-Jun | 1,171,928 | 116 | 9.04 | 24.0 | 15.0 |
| 2016 | 17-Feb | 30-Jun | 702,028 | 135 | 7.73 | 21.0 | 13.2 |
| 2015 | 10-Mar | 28-Jun | 957,706 | 111 | | | |
| 2014 | 5-Mar | 30-Jun | 608,661 | 118 | | | |
| 2013 | 4-Mar | 30-Jun | 993,089 | 119 | | | |
| 2012 | 13-Feb | 29-Jun | 590,157 | 138 | | | |
| 2011 | 24-Feb | 30-Jun | 841,043 | 127 | | | |
| 2010 | 25-Feb | 30-Jun | 471,415 | 126 | | | |
| 2009 | 12-Feb | 30-Jun | 2,174,478 | 139 | | | |
| 2008 | 13-Feb | 28-Jun | 2,695,955 | 137 | | | |
| 2007 | 16-Feb | 30-Jun | 785,887 | 135 | | | |
| 2006 | 20-Feb | 29-Jun | 130,024 | 130 | | | |
| 2005 | 26-Feb | 29-Jun | 70,157 | 124 | | | |
| 2004 | 8-Feb | 29-Jun | 315,018 | 143 | | | |
| 2003 | 12-Feb | 16-Jun | 437,693 | 125 | | | |

### 3.2.3 Results and discussion

We formally set $\Delta t = T_{\text{emp}} \times 2 \cdot 10^{-5}$, which is sufficiently small compared with $T_{\text{emp}}$, as verified in the previous study [38], while in the actual computation, we normalize the governing equations by $T_{\text{emp}}$ and $S = S_{\text{emp}}$, and after computation, we denormalized each quantity. In this section, time 0 is the beginning of February 1.

**Figure 3** shows a set of sample paths of the WT $w$, biological clock $\tau$, and unit-time fish count $X$, where we set $\omega = 2$ and assume stochastic WT dynamics (22). The sample path of WT $w$ gradually increases with fluctuations, while that of the unit-time fish count $X$ is intermittent, having bursts and rests, which has been found to be a characteristic of the upstream migration of Ayu [35]. In this sample path, migration starts and ends at 26.4 (day) and 148.5 (day), respectively, with a migration duration of 122.1 (day). The path of WT hits the upper boundary ($w_t = \bar{w}$) near the end of migration, after which $M_t = \omega$ is satisfied. **Figure 4** shows the computed histograms of the start date, end date, migration duration (except for the point mass at $T = T_{\text{emp}}$ with a weight of 47%), and total fish count of migration for the nominal case with 1,000,000 sample paths. All these histograms are unimodal with visible right tails. Note that the start day purely depends on the WT dynamics, while the end date and total fish count depend both on WT and unit-time fish count.

We perform sensitivity analysis of the proposed model, where we investigate the dependence of the unit-time fish count on the parameters $a^{(w)}$, $b^{(w)}$, $T_{\text{emp}}$, and $\omega$. The parameter values examined below have been determined based on a hypothesis that the start and end dates of the upstream migration of Ayu are within a reasonable range in the sense of average and standard deviation considering **Table 2**. The total number of sample paths generated is 1,000,000 for each computation, so we can rely on the first three digits of the statistics. **Tables B2-B6** in **Appendix B** summarize statistics about the start and end dates and the duration of migration corresponding to the computational cases examined below.

**Figure 5** shows histograms of the total fish count for the nominal case, deterministic case $a^{(w)} \to +\infty$, less-fluctuating case $b^{(w)} \to \sqrt{1/2}b^{(w)}$, and more fluctuating case $b^{(w)} \to \sqrt{2}b^{(w)}$ in the WT dynamics. This situation corresponds to doubling or halving the variance in the WT. The increase in $b^{(w)}$ increases the fluctuation in the migration duration but not in the mean, and at the same time, the start date becomes later with an approximately 4 (day) delay in the mean. Additionally, there is an approximately 5 (day) difference between the deterministic case $a^{(w)} \to +\infty$ and the more fluctuating case $b^{(w)} \to \sqrt{2}b^{(w)}$; the latter results in a shorter migration duration in the mean. On the other hand, regarding these fluctuation characteristics of WT dynamics, the corresponding unit-time fish count is not critically affected by either the mean or standard deviation, e.g., the relative difference in average total fish counts between the deterministic and more fluctuating cases is approximately 0.5%. The relative differences among the other cases are also small. This is considered due to the theoretical assumption about the coefficients $a, \sigma$ in the CIR bridge (**Appendix B**) that are almost vanishing around the start and end dates; indeed, the identified parameter values of $m, n, p, q$ are significantly larger than 1, as shown in **Table 1**. In other words, the regions where the average is small serve as buffer zones against distortions in environmental factors, which is WT in our case, as demonstrated in **Table B5**. This finding is consistent with a recent report that fish migration is not always controlled by local environmental factors but by calendar dates [16]. Since our model was not constructed to reflect their research findings, there is no direct connection between the two. Due to its structure, this model alone cannot reveal the specific mechanisms underlying these calendar-dependent biological behaviors; however, at least in their case studies, simply increasing the number of factors in a model is not always the best approach.

We also compare the models with $\omega = 1, 2, 10, \text{ and } 100$. The total fish count seems not to be critically affected by the choice of $\omega$, as in the case of the WT dynamics discussed above. This is also considered due to the robustness of the proposed model, which is due to the condition $m, n, p, q >> 1$. This means that the same may not hold true for other fish species such that $m, n, p, q$ is closer to or smaller than 1. In summary, the influence of $\omega$ is negligible for the present case if its value is close to 1. Although it is purely a result of the findings, $\omega$-dependence of the upstream migration of Ayu seems to be small.

Finally, **Figure 6** shows histograms of the total fish count for the nominal $T_{\text{emp}}$ value and modulated values with replacement $T_{\text{emp}} \to (1 \pm 0.1) T_{\text{emp}}$ and randomized $T_{\text{emp}}$ in the CIR bridge. In view of **Remark 2**, these situations correspond to hypothesizing some genetic difference among years that triggers the difference in the threshold value $T_2$; according to the real data, the migration duration may vary approximately $\pm 10$ (day) from the mean, which approximately corresponds to $\pm 10\,\%$ in $T_{\text{emp}}$ in the present case. The stock of larvae in the sea [58] may also affect $T_{\text{emp}}$; it would become larger as there exists a larger amount of stock. Here, the randomization of $T_{\text{emp}}$ is carried out so that it follows a uniform distribution, which is the simplest distribution that we can assume here, with a mean of 127.8 (day) and standard deviation of 8.8 (day), which are consistent with **Table 2**. These computational results along with **Table B4** show that the influences of $T_{\text{emp}}$ are significant for our case because they directly determine the migration duration. The 10% increase and decrease in $T_{\text{emp}}$ yields approximately 10% increase and decrease in the average of the total fish count, respectively. We found a similar tendency for the standard deviation of the total fish count. The randomized $T_{\text{emp}}$ case results in larger standard deviations in the end date and duration of migration. These findings imply that, for the Ayu case, estimation of the migration duration $T_{\text{emp}}$ plays an important role in modeling upstream fish migration phenomena.

Finally, we examine the cases with accelerating and slowing-down average temperature increases over time: $\hat{\kappa} \to (1 \pm 0.1) \hat{\kappa}$, which would correspond to hypothetical climate change scenarios; warming scenarios seem to be more likely [59,60], but here we consider both warming and cooling scenarios. **Table B6** shows that modulating $\hat{\kappa}$ in this range does not critically affect the total fish count but rather the migration duration. Approximately 5 (day) forward and delay of the start date with approximately 2-3 (day) shortened and extension of the migration duration, respectively, for the cases $\hat{\kappa} \to 1.1\hat{\kappa}$ and $\hat{\kappa} \to 0.9\hat{\kappa}$; however, the total fish count seems not to heavily depend on these parameter changes. This finding suggests that the schedule of surveys of upstream migration of Ayu, which is conducted annually in many rivers in Japan, should be adjusted or its duration adapted to take climate change trends into account.

If similar data sets about fish counts are available in other regions, then the proposed model may be applicable there as well; however, as pointed out in Yoshioka [37], the reliability of manual observations, especially those conducted only for a limited time each day, is not necessarily high. It should be noted that, in the case of the Nagara River, the existence of a fully automated observation system made the above modeling possible. We hope that the widespread adoption of such a reliable data collection method will deepen the understanding of fish migration.

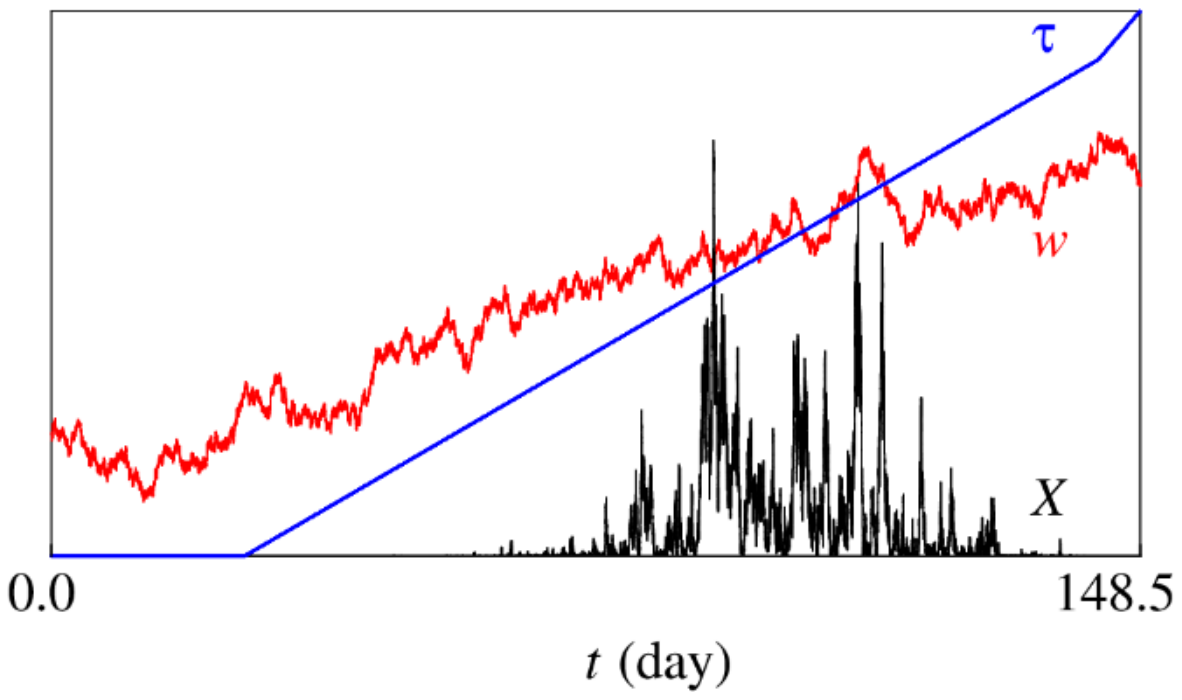

**Figure 3.** A set of sample paths of WT $w$ (°C), biological clock $\tau$ (day), and unit-time fish count $X$ (ind/day) with normalization for visualization proposals. Note that this figure is intended to help visualize the qualitative behavior of each variable.

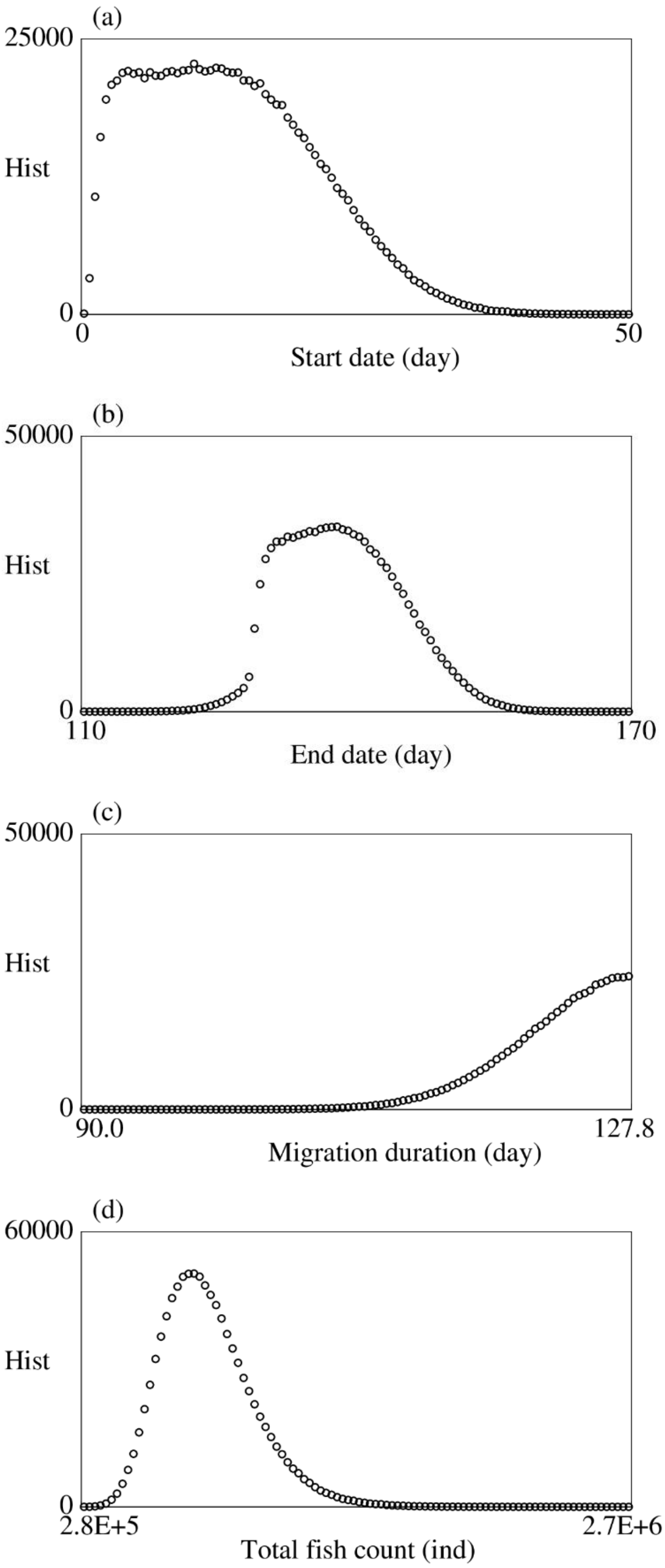

**Figure 4.** Histograms (Hist) of (a) start date (day), (b) end date (day), (c) migration duration (except for the 47% mass at 127.8), and (d) total fish count (ind) for the nominal case.

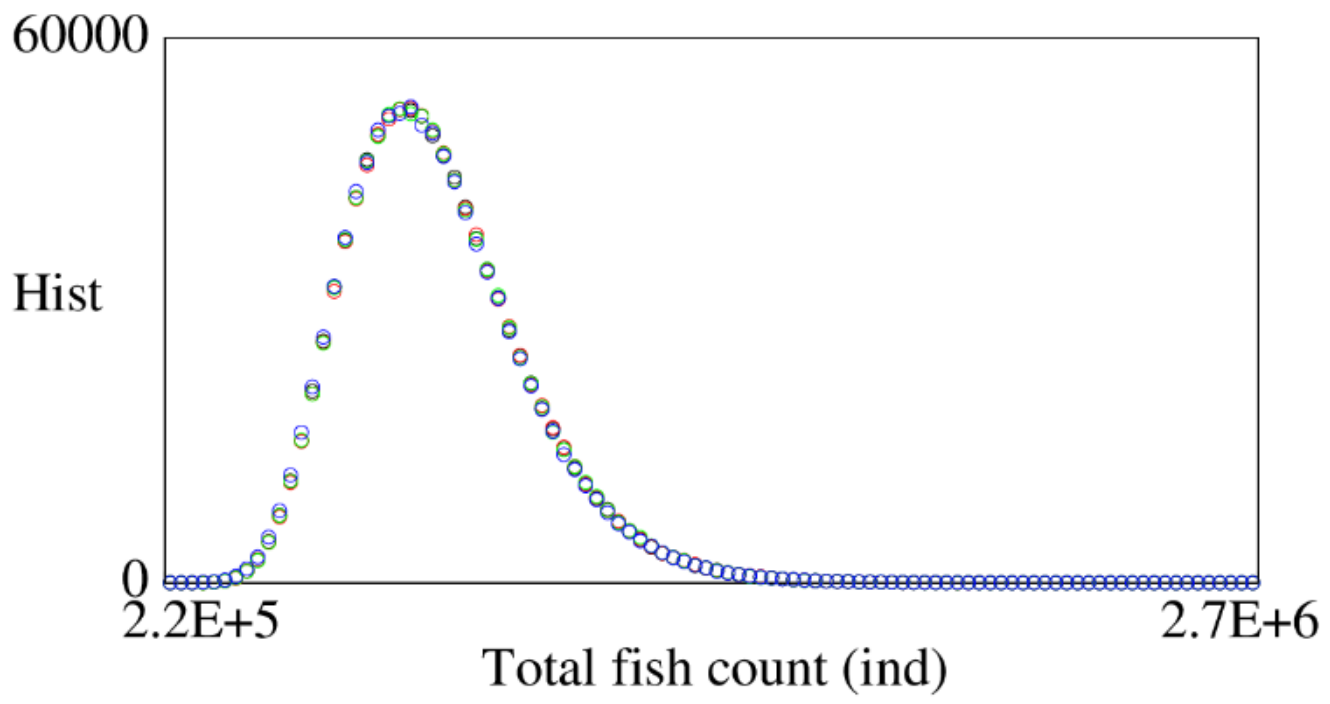

**Figure 5.** Histograms of the total fish count (ind) for the nominal case (black), deterministic case $a^{(w)} \to +\infty$ (red), less-fluctuating case $b^{(w)} \to \sqrt{1/2}b^{(w)}$ (green), and more fluctuating case $b^{(w)} \to \sqrt{2}b^{(w)}$ (blue).

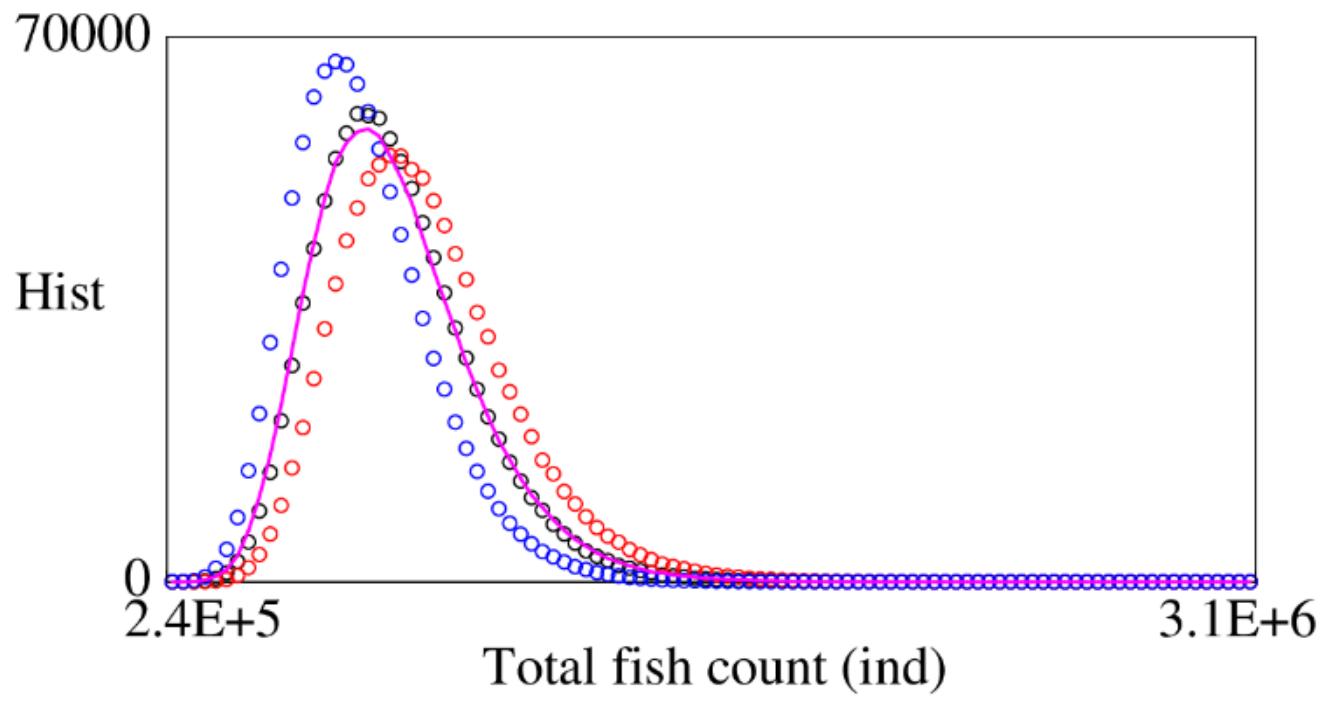

**Figure 6.** Histograms of the total fish count (ind) for the nominal case (black), $T_{\mathrm{emp}} \to 1.1T_{\mathrm{emp}}$ (red), $T_{\mathrm{emp}} \to 0.9T_{\mathrm{emp}}$ (blue), and randomized $T_{\mathrm{emp}}$ (magenta).

### 3.3 Hii River study: heuristic eDNA modeling

#### 3.3.1 Site description and data

The second target of this study is the Hii River, a first-class river that flows through the Sanin region of Japan and finally empties into the Sea of Japan (**Figure 7**). The main study site is the Negai Bridge in the Kisuki area (called Kisuki in the sequel: 35°17'35"N, 132°53'51"E) at approximately 30 (km) upstream from Lake Shinji, which is a part of a brackish lake system downstream of the main branch.

In the Hii River, unlike in the Nagara River, automated observation of the migration of Ayu has not been conducted. Instead, with the help of the Hii River Fisheries Cooperative, the first author has been investigating migration dynamics based on weekly eDNA sampling (from April to November in 2025; the data quantification was due to Polymerase Chain Reaction (PCR) [61]) in conjunction with 20-min WT observations (from March 15 to August 21 in 2025; the measurement due to U20, Onset). To avoid a decrease in the accuracy of eDNA analysis, we avoided or rescheduled sampling dates when the sampled water was turbid. Ayu in the Hii River contributes to the formation of local social culture, as evidenced by events such as net-casting competitions organized by the Hii River Fisheries Cooperative. Note that we assumed that weekly sampling suffices for the eDNA modeling here, by assuming that the eDNA dynamics results from some averaging of fish migration dynamics. High-resolution and/or larger amount of sampling would be needed for implementing more advanced models.

Regarding the Hii River, studies have frequently been conducted on the structure and ecology of the connected brackish water system in its lower reaches (Ramsar Convention wetland) [62,63], as well as on flood control and environmental disturbances caused by floods [64,65]. While there is accumulated data on the growth of individual Ayu in the Hii River [66], studies related to their upstream migration dynamics, particularly modeling, have not been conducted thus far. This paper investigates the applicability of CIR bridges to eDNA data to deepen our understanding of Ayu in the Hii River.

In the sequel, the beginning of February 1, 2025, is set to the time 0. **Figure 8** shows the observed data of the eDNA of Ayu and daily WT at Kisuki. The daily WT is expected to increase on average from February to August in 2025. In 2025, a positive eDNA concentration value was first detected on April 17. The eDNA concentration has two visible peaks, where the former corresponds to the upstream migration of juvenile fish, while the latter corresponds to the downstream migration of adult fish. We need to somehow separate the two migration events, which is addressed based on the empirical analysis conducted in the Nagara River case in which the upstream migration ends around the daily WT of 23 (°C). The upstream migration has then been estimated to be terminated on July 14 in 2025, as shown in **Figure 8**. The positive eDNA concentration value after this date is considered because the eDNA comes from the upstream of Kisuki; it becomes positive if there is an Ayu population in a nearby upstream reach from the study site. We therefore focus on the eDNA data before this date.

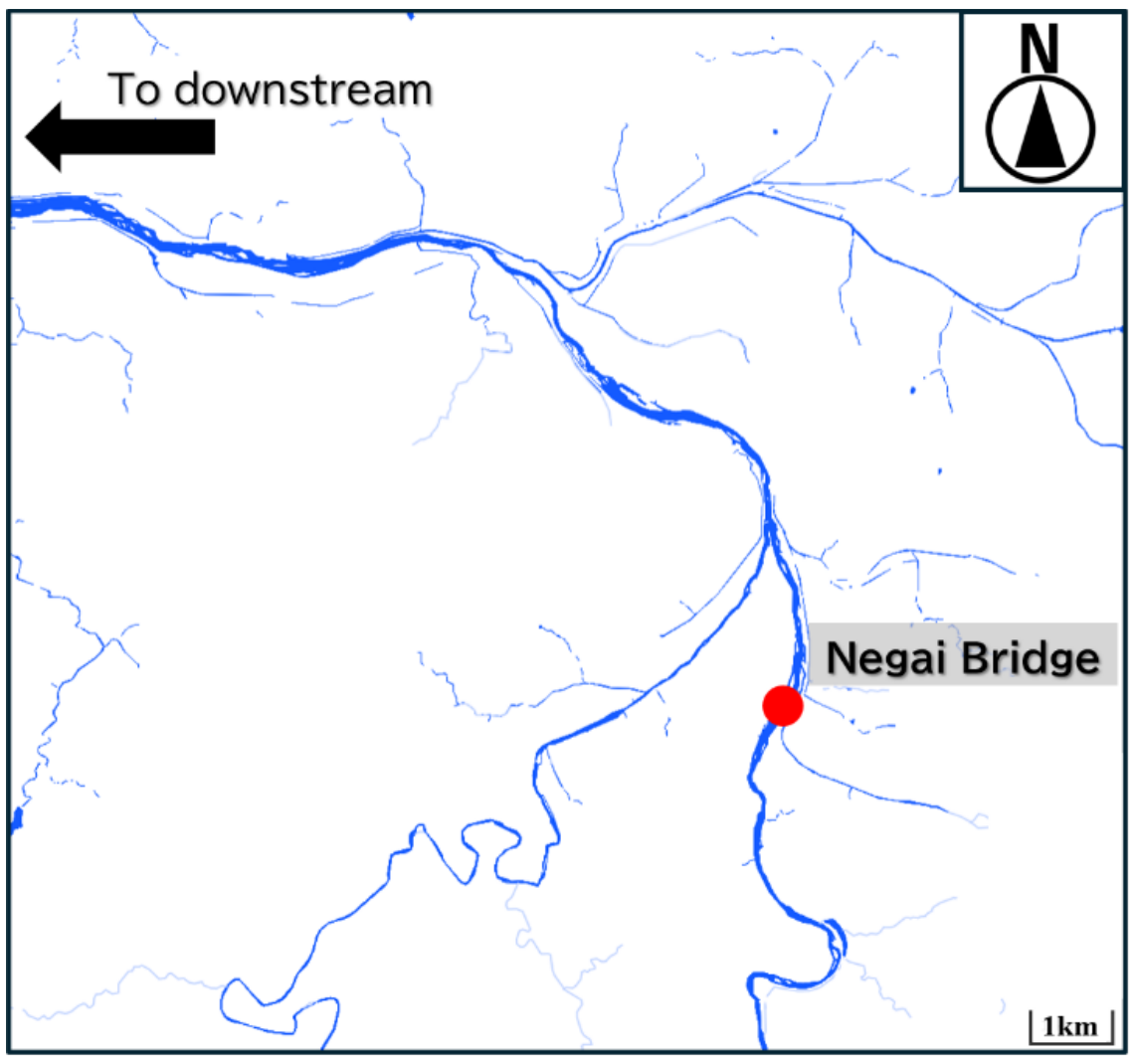

**Figure 7.** Map of the study site in the Hii River (modified from Chiri-in Chizu Vector: https://github.com/gsi-cyberjapan/gsimaps-vector-experiment).

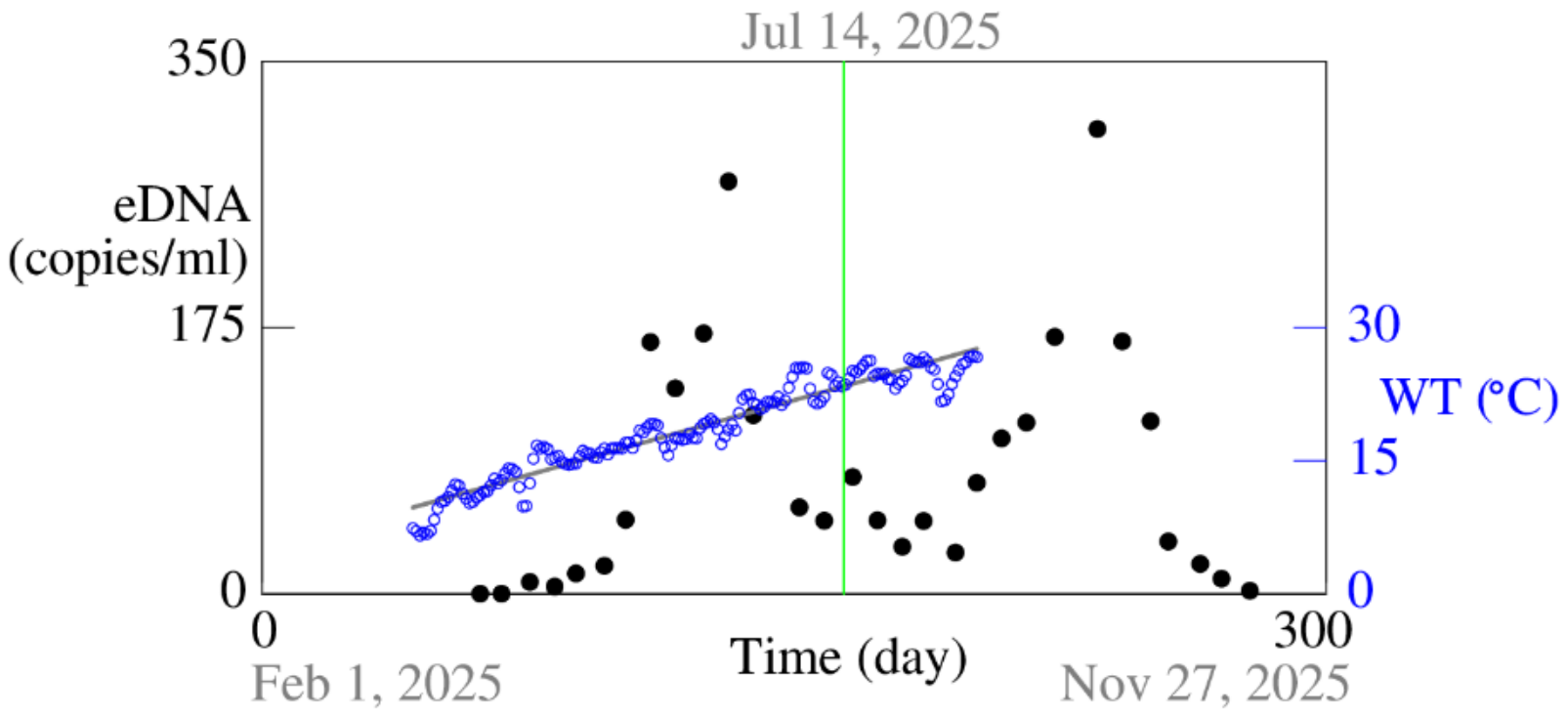

**Figure 8.** Observed data of the eDNA concentration (copies/ml) of Ayu (black circles) and daily WT (°C) (blue circles) at Kisuki. The gray line represents the linear regression of the daily WT data, and the green vertical line corresponds to July 14, around which WT exceeds 23 (°C).

### 3.3.2 An extended model

This application example aims to conduct an exploratory analysis of how CIR bridges can be used for analyzing environmental DNA data. We cannot directly apply CIR bridges to the eDNA data because it models unit-time fish count, while eDNA data are an aggregated index that quantifies fish populations located upstream of the observation point, i.e., fish that have finished their upstream migration through the observation point. In other words, eDNA only indirectly counts migrants. Additionally, eDNA degrades in natural water environments as time elapses [e.g., 67,68]. This implies that modeling eDNA requires another mechanism that describes aggregation and degradation.

In this paper, we propose the following model for eDNA time series at an observation point:

$$\mathrm{d}E_t = \left(GX_t^H - RE_t\right)\mathrm{d}t,\quad t > \theta_1 \tag{23}$$

subject to $E_{\theta_1} = 0$, where $E = \left(E_t\right)_{t\geq 0}$ (copies/ml) is the eDNA concentration in units of (copies/ml), $R > 0$ (1/day) is the attenuation constant for modeling the attenuation and degradation, $G > 0$ (copies/ml/day) and $H > 0$ (-) are allometric constants [e.g., 69,70], and $X_t$ is obtained from a CIR bridge. The differential equation (23) coupled with the CIR bridge (4) outputs the eDNA concentration. Note that the present submodel is used as a preliminary heuristic tool in this paper.

***Remark 6*** By **Proposition 1**, the pathwise unique existence of the process $E$ holds true, and thus the extended model here is well-posed.

### 3.3.3 Results and discussion

We set $\omega = 2$ following the previous application. The daily WT data have been reasonably regressed by a linear model, as shown in **Figure 8**, where the model has an intercept of $\kappa_0 = 4.885$ (°C) and a slope of $\kappa_1 = 0.1125$ (°C/day) ($R^2$ value: 0.917). The two parameters $\eta$ and $\lambda$ in the stochastic WT dynamics have also been identified as $\eta = 0.1420$ (1/day) and $\lambda = 0.7977$ (°C/day$^{1/2}$).

In this application, for the CIR bridge, we set $m = n = 10$ considering the first example, from which we found $m > n >> 1$ with $A = 10^6$ (-); this is due to the linearity of (23) where influences of $A$ can be absorbed into $G$. This setting is sufficient if the objective is to simulate the eDNA concentration only where $X$ plays the role of a latent variable. This means that the eDNA concentration behaves as a non-Markovian process because the information contained in a trajectory of $E$ is strictly less than that of the couple $(X, E)$.

Based on the WT analysis in the midstream Nagara River, we infer that upstream migration in the downstream section began on March 10. More specifically, we assume the same WT threshold values for $\overline{w}, \underline{w}$ as the Nagara River case, yielding that the upstream migration at Kisuki in 2025 was supposed to start on March 10 and end on July 14, and the resulting migration duration is $T_{\mathrm{emp}} = 127$ (day). For the middle reaches, the start time of upstream migration is assumed to be April 3, 2025—the date of the initial eDNA sampling, which yielded a value of "0". This assumption is based on the consideration that there must be a time lag (as discussed later) between the first confirmed migration in the midstream section and that in the downstream section; in the future, it would be desirable to simultaneously monitor water temperature and migration dynamics in both the downstream and midstream sections.

Given $A, m, n$, the values $V, p, q$ are determined based on the empirical finding that the coefficient of variation of daily fish count is close to a constant value, which is approximately 2 [38], yielding $p = 2m$, $q = 2n$, and $V = 4A^2$. The value of $r$ is also taken from the first application. The remaining parameters $G, H, R$ are determined by a least-squares method between the theoretical average values and empirical values of eDNA concentration, where the former is generated by numerically computing (23) with the linearization based on the delta method (Theorem 6.25 in Henze [71]) with the aforementioned ansatz $2\mathbb{E}[X_t] = \sqrt{\mathbb{E}\left[\left(X_t - \mathbb{E}[X_t]\right)^2\right]}$, which is again heuristic:

$$\mathbb{E}\left(X_t^H\right) \to \left(\mathbb{E}[X_t]\right)^H + \frac{1}{2}H(H-1)\left(\mathbb{E}[X_t]\right)^{H-2} \times \mathbb{E}\left[\left(X_t - \mathbb{E}[X_t]\right)^2\right] = \left(1 + 2H(H-1)\right)\left(\mathbb{E}[X_t]\right)^H . \quad (24)$$

This delta method can be considered a version of mean-field approximations that assumes eDNA to be a spatiotemporally mixed quantity. From a first-principles modeling perspective, approaches from local fluid dynamics and transport phenomena may be employed [e.g., 72,73], but it is unclear how meaningful these approaches are for conceptual models such as the proposed one.

We examine both linear ($H = 1$) and nonlinear cases ($H \neq 1$) as shown in **Table 3** with the visualizations in **Figure 9**, demonstrating that the nonlinear model 5% better captures the falling limb of the empirical eDNA concentration profile in view of the RMSE. The nonlinear model suggests $H \approx 0.75 < 1$, which is consistent with the allometric relationship between the population density and eDNA concentration of Ayu in a different river [69]; the eDNA concentration depends sublinearly on the fish population. The attenuation coefficient is approximately 0.4 (1/day) in both cases, suggesting that the memory of the fish count remains in the eDNA concentrations for a couple of days.

We also briefly conduct numerical computation with the same computational resolution as the first application. The simulation presented below shows the analysis results for the downstream reach of the Hii River, based on the assumption of a 25-day lag (corresponding to the period between March 10 (migration started at downstream) and April 3 (migrants reached midstream)) in the upstream migration dynamics of Ayu between the downstream and midstream (Kisuki). More specifically, here the interval $(\theta_1, \theta_2)$ corresponds to the migration dynamics in the downstream reach, whereas the shifted interval $(\theta_1 + 25, \theta_2 + 25)$ corresponds to the upstream migration dynamics in the midstream reach. The specified lag, although it is heuristic, is reasonable considering that Ayu migrates upstream around 0.3-0.5 to a couple of days[4] and that the Negai bridge is approximately 30 (km) upstream from Lake Shinji. The computational results presented below therefore concern the fish migration in the downstream (that for Kiski can be obtained by simply shifting 25 (day) by the aforementioned assumption).

[4] Takahashi Research Office of Freshwater Biology: https://hito-ayu.net/introduction01.html. Last accessed on August 13, 2026.

For the nonlinear model driven by the stochastic WT dynamics, **Figure 10** shows sample paths of the (nondimensionalized) unit-time fish count $X$ and eDNA concentration $E$, demonstrating that the profile of the eDNA concentration is indeed seen as a smoothed and scaled one of the unit-time fish counts. **Figure 11** compares the average (Ave) and standard deviation (Std) between the linear and nonlinear models, showing that the nonlinear model predicts a larger average than the linear ones, which is partly due to the model fitting procedure based on a mean-field ansatz and a slightly large variance, while both models give comparable standard deviation profiles. Here, what is important is that both models qualitatively give the same results and hence are considered not easy to distinguish from data; we need to accumulate knowledge from another standpoint to experimentally quantify the allometric relationship [74].

In the future, mathematical modeling, particularly focusing on the magnitude of the coefficients, will be necessary. For example, to investigate whether the result of $H$ is coincidental, we are currently conducting continuous data collection at multiple locations along the Hii River. The transferability of the models can also be another issue. In this study, we have also applied the model with parameter values $m,n$ identified for the Nagara River; however, we ended up with the attenuation constant $R$ being nearly 100, indicating that the attenuation is not working effectively in this case. This situation is unphysical. Therefore, directly transferring a model from one river to another would need care. In contrast, the order estimate of parameters such as that utilized in this paper would be helpful for application case studies. Although the results obtained in this application example are limited, as discussed above, they are considered to suggest the potential application of CIR bridges to eDNA analysis.

**Table 3.** Fitted parameter values and root-mean square errors (RMSEs).

| Parameter | Linear case | Nonlinear case |
|---|---|---|
| $H$ (-) | 1 | 7.476E-01 |
| $G$ (copies/ml/day) | 9.533E+01 | 1.369E+02 |
| $R$ (1/day) | 4.112E-01 | 3.954E-01 |
| RMSE | 3.174E+1 | 3.025E+1 |

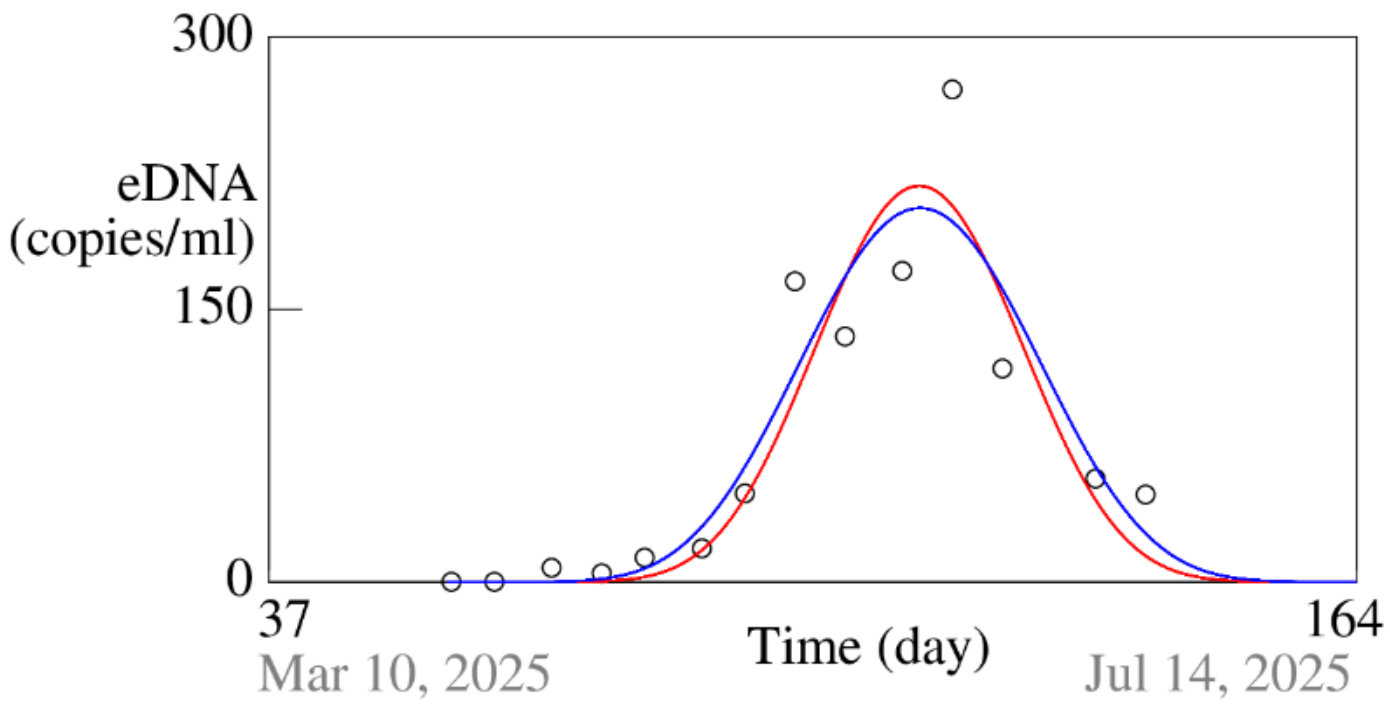

**Figure 9.** eDNA concentration (black) (copies/ml) For the empirical result (black) and theoretical ones for the linear (red) and nonlinear cases (blue) in the midstream Hii River.

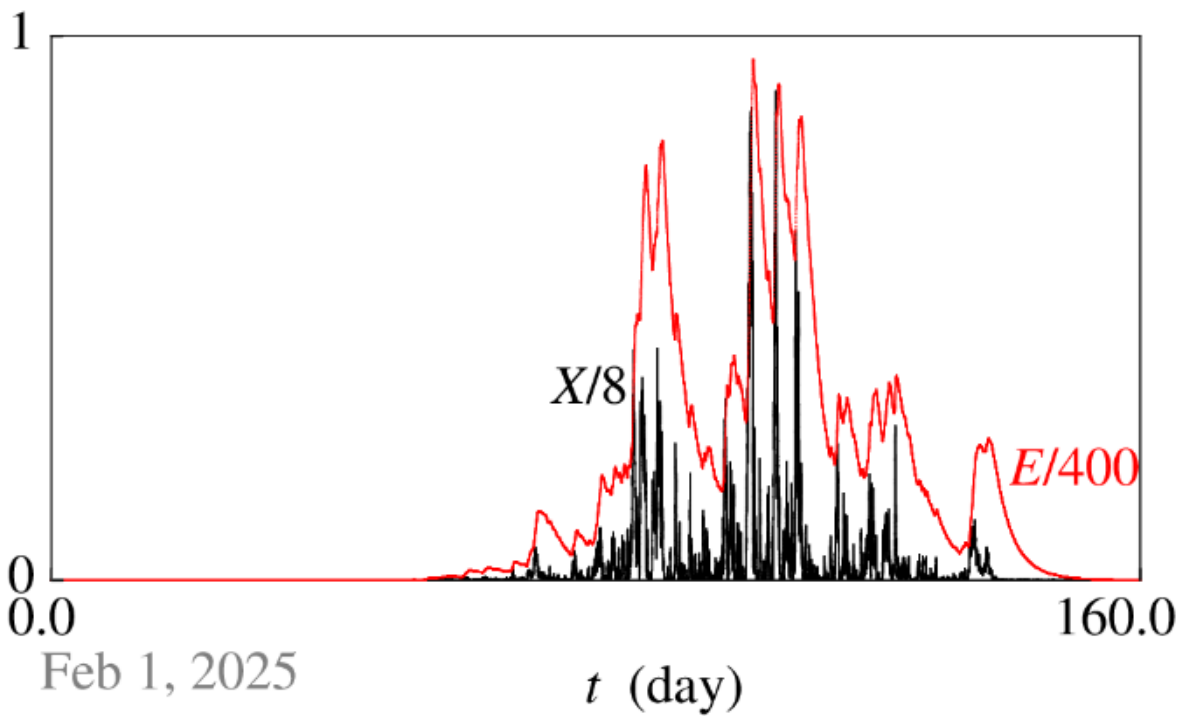

**Figure 10.** Sample paths of the (nondimensionalized) unit-time fish count $X$ (-) and eDNA concentration $E$ (copies/ml) in the downstream Hii River.

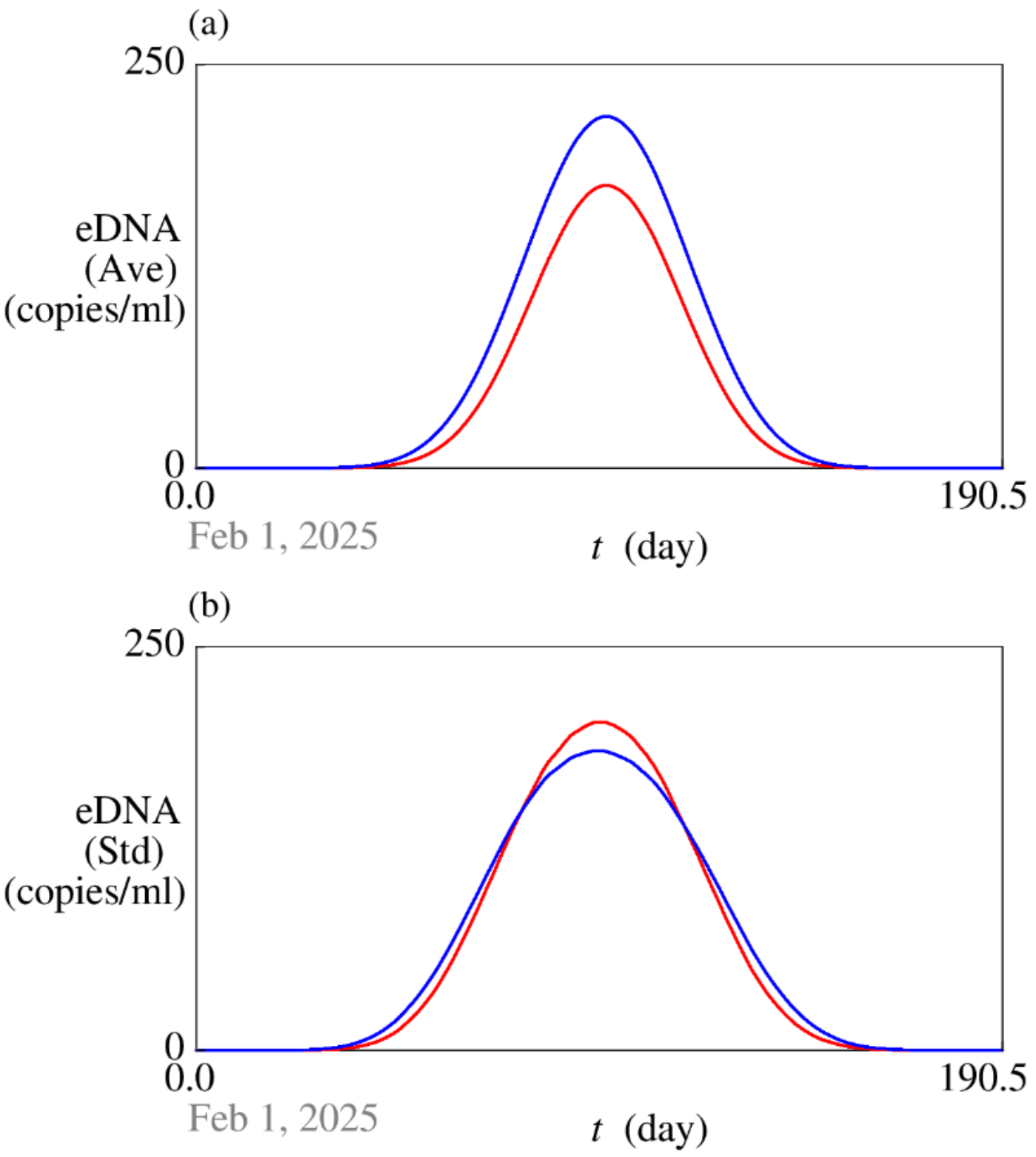

**Figure 11.** Comparison between the average (Ave) (copies/ml) and standard deviation (Std) (copies/ml) of the eDNA concentration between the linear (red) and nonlinear models (blue) in the downstream Hii River.

## 4. Conclusion

By focusing on fish migration phenomena, this paper proposed a novel CIR bridge with randomized initial and terminal times, which are determined as stopping times of an integration of a nonnegative and increasing process. These stopping times represent the initiation and termination of seasonal fish migration observed at a fixed location. The well-posedness of the proposed CIR bridge was established under boundedness conditions of the stopping times, which are considered natural for applications to fish migration. This paper also applied the CIR bridge to the upstream migration of juvenile Ayu in rivers in Japan. We estimated parameter values of the model based on a hypothesis that WT is a dominant factor that influences the migration of the fish. We also conducted computational analysis of the model to investigate its performance and parameter sensitivity. We then discussed the implications of the proposed CIR bridge for eDNA data from Ayu collected in another river environment to gain a deeper mathematical understanding of the migration of this fish species.

For highly fluctuating subjects, such as fish count or environmental DNA data, an SDE-based approach is considered effective. This study is a step forward in the mathematical modeling of fish migration phenomena, but there still exist limitations to be overcome. For example, we only dealt with the upstream migration of a single fish species and its downstream migration, and moreover, whole life-history modeling remains unresolved. While it may be possible to construct complex models to simulate how environmental factors other than WT affect fish migration, the usefulness of a mathematically simple model cannot be denied. We believe it is necessary to advance theory building while maintaining a balance between the complexity and usability of the model. Incorporating genetic information from individual fish might improve the model, but collecting such data requires considerable research effort. Also, in this paper, although it is purely a result of the findings, the $\omega$-dependent nature of the migration of ayu was small; what about other fish species? WT-dependence of the fish migration can also be improved by further accumulating the biological data about Ayu, such as otolith and capture data.

We focused on a specific diffusion bridge, but it will be possible to randomize the initial and final times in other models, such as Brownian bridges and other SDE-based models, using a similar framework. Finally, through collaboration between the author, collaborators, and fisheries cooperatives, eDNA analysis and growth surveys of Ayu in the Hii River are currently underway. We will combine these data with the results obtained to develop an integrated model for the management of inland fisheries resources in the future.

## Appendix

### A. Proofs

***Proof of Proposition 1***

First, we have $\theta_1 < \theta_2$ due to **Assumption 1**. Second, we prove that the SDE (6) admits a pathwise unique solution in $(T_1, T_2)$ that is nonnegative and satisfies $Y_{T_1} = Y_{T_2} = 0$ a.s.. If this claim is proven, then we can conclude this proof because the process $\tau$ is continuous and strictly increasing for $t > 0$, and therefore, the transformation of variables $s = \tau_t$ is well-defined. Furthermore, **Assumption 1** leads to that $\tau_t$ is strictly increasing in $(\theta_1, \theta_2)$, and $\theta_1 < \theta_2 < +\infty$ a.s. due to $T_1 < T_2 < +\infty$ and $\int_0^\infty M_t \mathrm{d}t = +\infty$. These observations imply that the domain $(\theta_1, \theta_2)$ is not null and that the process $X$ is nonnegative and satisfies the desired conditions $X_{\theta_1} = X_{\theta_2} = 0$. Moreover, then the SDE (4) is truly a time change of the SDE (6), and therefore, the desired unique existence result for the former SDE holds true if we can ensure the existence of $\hat{B} = (\hat{B}_s)_{s \geq 0}$.

We give an explicit construction of the Brownian motion $\hat{B}$. We define $\hat{B}$ as

$$\hat{B}_s = \int_0^{\tau^{-1}(s)} \sqrt{M_u} \mathrm{d}B_u . \tag{25}$$

This construction makes explicit why the strict positivity $M_t > 0$ a.s. is essential for the time change. Under the time change $s = \tau_t = \int_0^t M_u \mathrm{d}u$, the diffusion term of SDE (4) transforms as follows:

$$\sigma_t \sqrt{M_t X_t} \mathrm{d}B_t = \sigma_t \sqrt{M_t Y_s} \mathrm{d}B_t = \hat{\sigma}_s \sqrt{Y_s} \cdot \sqrt{M_t} \mathrm{d}B_t , \tag{26}$$

where $\hat{\sigma}_s = \sigma_t$. For this to match the form $\hat{\sigma}_s \sqrt{Y_s} \mathrm{d}\hat{B}_s$, the process $\hat{B}$ must satisfy the relationship $\mathrm{d}\hat{B}_s = \sqrt{M_t} \mathrm{d}B_t$. The natural candidate of $\hat{B}$ is therefore given by (26). The process $\hat{B}$ is a continuous local martingale with $\hat{B}_0 = 0$. To confirm it is a standard Brownian motion, we compute its quadratic variation. By the Itô isometry, we have ($\langle \hat{B} \rangle_s$ represents the quadratic variation of $\hat{B}$ at time $s$)

$$\langle \hat{B} \rangle_s = \int_0^{\tau^{-1}(s)} \left(\sqrt{M_u}\right)^2 \mathrm{d}u = \int_0^{\tau^{-1}(s)} M_u \mathrm{d}u = \tau\left(\tau^{-1}(s)\right) = s , \tag{27}$$

where the last equality is due to the identity $\tau\left(\tau^{-1}(s)\right) = s$, which holds whenever $\tau_t$ is strictly increasing. By Lévy's characterization theorem (e.g., Theorem 3.3.16 in Karatzas-Shreve [48]), a continuous local martingale starting at 0 with quadratic variation equal to s is a standard Brownian motion. Hence $\hat{B}$ is indeed a standard 1-D Brownian motion. The identity $\tau\left(\tau^{-1}(s)\right) = s$ requires $\tau_t$ to be strictly increasing, which holds if and only if $M_t > 0$ Lebesgue a.e. a.s.. This shows that the strict positivity of $M$ imposed in **Assumption 1** is necessary not only for the invertibility of the time change in the drift, but also to ensure that the transformed diffusion noise $\hat{B}$ is a genuine standard Brownian motion.

Now, we prove the unique existence of nonnegative solutions to the SDE (6) in $(T_1, T_2)$ that is a.s. nonnegative and satisfies $Y_{T_1} = Y_{T_2} = 0$. To simplify notations, we can set $T_1 = 0$, $T_2 = 1$ without any loss of generality; this is simply a linear transformation using the parameter $s$. The proof here is based on Proof of Proposition 1 in Yoshioka [35], but we need a more careful estimate at the terminal time $s = 1$ so that there is a possible blow-up of the coefficients $\hat{a}, \hat{\sigma}$. By **Assumption 2**, there exist positive constants $\underline{r}, \overline{r}$ such that $\underline{r} \leq \hat{r}_s \leq \overline{r}$ ($0 \leq s \leq 1$). Then, Proof of Proposition 1 (the first paragraph) in Yoshioka [35] shows that the SDE (6) admits a unique pathwise solution if its domain is restricted to $(0, 1-\varepsilon)$ with any constant $\varepsilon \in (0,1)$, and the solution formally satisfies the following variation of constant formula:

$$Y_s = \underbrace{\int_0^s \hat{a}_u \exp\left(-\int_u^s \frac{\hat{r}_v}{1-v} \mathrm{d}v\right) \mathrm{d}u}_{N_s^{(1)}} + \underbrace{\int_0^s \hat{\sigma}_u \exp\left(-\int_u^s \frac{\hat{r}_v}{1-v} \mathrm{d}v\right) \sqrt{Y_u} \mathrm{d}\hat{B}_u}_{N_s^{(2)}}, \quad s \in [0, 1-\varepsilon]. \tag{28}$$

Here, the first term $N_s^{(1)}$ is simply a function of time, and the second term $N_s^{(2)}$ is a fluctuation defend at least in $[0, 1-\varepsilon]$. We have the average $\mathbb{E}[Y_s] = N_s^{(1)}$ and variance $\mathbb{V}[Y_s] = \mathbb{E}\left[\left(N_s^{(2)}\right)^2\right]$ in $[0, 1-\varepsilon]$. To see that the terminal condition $Y_1 = 0$ is satisfied with probability 1, in view of Lemma 1 in Yoshioka [75], it suffices to show that

$$\mathbb{E}[Y_s] \to 0 \text{ and } \mathbb{E}\left[\left(N_s^{(2)}\right)^2\right] \to 0 \text{ as } s \nearrow 1. \tag{29}$$

Below, we prove (29).

For the first limit (29), the boundedness assumption about $\hat{r}$ shows that, for any $\varepsilon \in (0,1)$ and $s \in [0, 1-\varepsilon]$,

$$\begin{aligned} N_s^{(1)} &= \int_0^s \hat{a}_u \exp\left(-\int_u^s \frac{\hat{r}_v}{1-v} \mathrm{d}v\right) \mathrm{d}u \\ &\leq \int_0^s \hat{a}_u \exp\left(-\int_u^s \frac{\underline{r}}{1-v} \mathrm{d}v\right) \mathrm{d}u \\ &= \int_0^s \hat{a}_u \left(\frac{1-s}{1-u}\right)^{\underline{r}} \mathrm{d}u \\ &= (1-s)^{\underline{r}} \int_0^s \hat{a}_u (1-u)^{-\underline{r}} \mathrm{d}u. \end{aligned} \tag{30}$$

The integrand in the last line is nonnegative and is on the order of $O\left((1-u)^{\alpha-\underline{r}}\right)$ near $u = 1$; therefore, we obtain the first limit in (29) if $\underline{r} + \alpha - \underline{r} + 1 > 0$, i.e., $\alpha > -1$, which holds true by **Assumption 1**; here, we temporally assumed $\underline{r} \neq 1$ for simplicity, but the case $\underline{r} = 1$ can be addressed in a similar way.

For the second limit (29), again, the boundedness assumption about $\hat{r}$ shows that, for any $\varepsilon \in (0,1)$ and $s \in [0, 1-\varepsilon]$, we have

$$
\begin{aligned}
\mathbb{E}\left[\left(N_s^{(2)}\right)^2\right] &= \int_0^s \hat{\sigma}_u^2 \mathbb{E}[Y_u] \exp\left(-\int_u^s \frac{2\hat{r}_v}{1-v}\mathrm{d}v\right)\mathrm{d}u \\
&\leq \int_0^s \hat{\sigma}_u^2 \mathbb{E}[Y_u] \exp\left(-\int_u^s \frac{2\underline{r}}{1-v}\mathrm{d}v\right)\mathrm{d}u \\
&= (1-s)^{2\underline{r}} \int_0^s \hat{\sigma}_u^2 (1-u)^{-2\underline{r}} \mathbb{E}[Y_u] \mathrm{d}u.
\end{aligned} \tag{31}
$$

Substituting (30) into (31) yields

$$
\begin{aligned}
\mathbb{E}\left[\left(N_s^{(2)}\right)^2\right] &\leq (1-s)^{2\underline{r}} \int_0^s \hat{\sigma}_u^2 (1-u)^{-2\underline{r}} (1-u)^{\underline{r}} \int_0^u \hat{a}_v (1-v)^{-\underline{r}} \mathrm{d}v\mathrm{d}u \\
&= (1-s)^{2\underline{r}} \int_0^s \hat{\sigma}_u^2 (1-u)^{-\underline{r}} \left(\int_0^u \hat{a}_v (1-v)^{-\underline{r}} \mathrm{d}v\right)\mathrm{d}u.
\end{aligned} \tag{32}
$$

The last integrand with respect to $s$ is nonnegative and is on the order of

$$
O\left((1-u)^{2\beta-\underline{r}+\alpha-\underline{r}+1}\right) = O\left((1-u)^{2\beta-2\underline{r}+\alpha+1}\right) \tag{33}
$$

near $u=1$. Therefore, we obtain the second limit in (29) if $2\beta-2\underline{r}+\alpha+1+2\underline{r}+1>0$, i.e., $2\beta+\alpha+2>0$, which holds true by **Assumption 1**. The proof is completed because we proved (29); here, we temporally assumed $\alpha-\underline{r}\neq -1$ and $2\beta-2\underline{r}+\alpha+1\neq -1$ for simplicity, but the other cases can be addressed in a similar way.

□

## B. Supporting results and data

### B.1 The baseline model

This section explains the CIR bridge with a fixed migration duration $[0,T]$ identified in Yoshioka [39]. This CIR bridge is just the original one (1), with a formal scaling of the last term satisfy (notations are the same as that in Yoshioka [39]):

$$
\mathrm{d}X_t = \left(a_t - \frac{r}{T-t}X_t\right)\mathrm{d}t + \sigma_t\sqrt{\frac{r}{T-t}X_t}\mathrm{d}B_t, \quad 0<t<T, \tag{34}
$$

where the coefficient $r>0$ is a constant and $a_t, \sigma_t$ are time-dependent curves:

$$
a_t = \frac{\mathrm{d}}{\mathrm{d}t}\mathbb{E}[X_t] + \frac{r}{T-t}\mathbb{E}[X_t] \tag{35}
$$

and

$$
\sigma_t^2 = \frac{T-t}{r}\frac{1}{\mathbb{E}[X_t]}\left(\frac{\mathrm{d}}{\mathrm{d}t}\mathbb{V}[X_t] + \frac{2r}{T-t}\mathbb{V}[X_t]\right). \tag{36}
$$

These equations are in fact the ordinary differential equations to determine the time-dependent average $\mathbb{E}[X_t]$ and variance $\mathbb{V}[X_t]$. In Yoshioka [39], these statistics were assumed to be given as parameterized curves as follows:

$$
\mathbb{E}[X_t] = A\left(\frac{t}{T}\right)^m\left(1-\frac{t}{T}\right)^n \tag{37}
$$

and

$$\mathbb{V}[X_t]=V\left(\frac{t}{T}\right)^p\left(1-\frac{t}{T}\right)^q \tag{38}$$

with positive parameters $A,V,m,n,p,q$ . These parameters have been identified by using a set of least-squares methods to the average, variance, and autocorrelation of $X$ , as detailed in Section 3.2 in Yoshioka [39]. In the identification procedure, for each year, the time was normalized so that $T\to 1$ and the unit-time fish count $X\to XT/S$ , where $S$ is the total number of annual migrants. With this normalization procedure, an empirical sample path of the daily unit-time fish count was considered as a discretized sample path of the normalized CIR bridge.

**B.2 Water temperature dynamics**

The parameters $a^{(w)}$ and $b^{(w)}$ in the SDE (22) for the WT dynamics at Oyabu were estimated as follows, which correspond to the inverse of the timescale and noise strength of the daily WT. The goal here is to estimate the values of these parameters to extend the deterministic model (21) to the SDE (22).

First, a year is fixed from 2016 to 2025. We have daily WTs $w_{j\Delta}$ ( $j=1,2,3,...,J$ ) from February 1 (day 1) to June 30 ( $J=151$ (day) for leap years and $J=150$ (day) otherwise) with the time increment $\Delta=1$ (day). The SDE to be estimated for this year is as follows:

$$\mathrm{d}w_t=-\eta\left(w_t-\left(\kappa_0+\kappa_1 t\right)\right)\mathrm{d}t+\lambda\mathrm{d}L_t,\ \ t>0 \tag{39}$$

with some real initial condition, where year-dependent constants $\eta,\lambda,\kappa_0,\kappa_1$ with $\lambda>0$ and $L=\left(L_t\right)_{t\geq 0}$ are a 1-D standard Brownian motion.

We follow the simple estimation method for OU processes. With the record $w_{j\Delta}$ ( $j=1,2,3,...,J$ ) of WTs in the chosen year, we estimate $\left(\kappa_0,\kappa_1\right)$ as the minimizer of the following least-squares error:

$$\sum_{j=1}^{J-1}\left(w_{(j+1)\Delta}-w_{j\Delta}+\eta\left(w_{j\Delta}-\left(\kappa_0+\kappa_1 j\Delta\right)\right)\Delta\right)^2 . \tag{40}$$

Second, having estimated $\left(\kappa_0,\kappa_1\right)$, we estimate $\eta$ as follows, which is the linear regression between $w_{(j+1)\Delta}-w_{j\Delta}$ and $\kappa_0+\kappa_1 j\Delta-w_{j\Delta}$ by using a line passing through the origin:

$$\eta=\frac{\sum_{j=1}^{J-1}\left(w_{(j+1)\Delta}-w_{j\Delta}\right)\left(\kappa_0+\kappa_1 j\Delta-w_{j\Delta}\right)\Delta}{\sum_{j=1}^{J-1}\left(\left(\kappa_0+\kappa_1 j\Delta-w_{j\Delta}\right)\Delta\right)^2} . \tag{41}$$

Third, having estimated $\left(\kappa_0,\kappa_1,\eta\right)$ , we estimate $\lambda$ as follows considering that the variance of the residual $w_{(j+1)\Delta}-w_{j\Delta}+\eta\left(w_{j\Delta}-\left(\kappa_0+\kappa_1 j\Delta\right)\right)\Delta$ should equal $\lambda^2\Delta$ :

$$\lambda^2 = \sum_{j=1}^{J-1} \left( w_{(j+1)\Delta} - w_{j\Delta} + \eta \left( w_{j\Delta} - \left( \kappa_0 + \kappa_1 j\Delta \right) \right) \Delta \right)^2 , \tag{42}$$

and then we use $\lambda > 0$. **Table B1** shows the estimated parameter values $(\kappa_0, \kappa_1, \eta, \lambda)$ for each year. Finally, we estimate $a^{(w)}$ and $b^{(w)}$ in the main text as the yearly average values of $\eta$ and $\lambda$, respectively, i.e., the last line in **Table B1**. The yearly average of the interception $\kappa_1$ (0.112 (°C/day)) is close to the slope $\hat{\kappa}$ of the deterministic model (21) (0.111 (°C/day)), demonstrating their consistency. For visualizations of the regression each year, see **Figure B1**.

**B.3 Supplementary analysis results**

First, **Table B2** compares the average (Ave) and standard deviation (Std) of the start date, end date, migration duration, and total fish count for the nominal case, deterministic case $a^{(w)} \to +\infty$, less fluctuating WT case $b^{(w)} \to \sqrt{1/2} b^{(w)}$, and more fluctuating WT case $b^{(w)} \to \sqrt{2} b^{(w)}$. **Table B3** compares the results for different values of $\omega$. **Table B4** compares the results for different values of $T_{\mathrm{emp}}$.

Second, we conduct sensitivity analysis against the parameters $m, n$ that determine the average curve (37). To maintain the theoretical consistency between the nominal and modified models, based on the relationship (20) about the total fish count, the new parameter values of $m, n$, which are denoted here by $m', n'$, are determined so that the following equality is satisfied:

$$A \int_0^{T_{\mathrm{emp}}} \left( \frac{t}{T_{\mathrm{emp}}} \right)^m \left( 1 - \frac{t}{T_{\mathrm{emp}}} \right)^n \mathrm{d}\left( \frac{t}{T_{\mathrm{emp}}} \right) = A' \int_0^{T_{\mathrm{emp}}} \left( \frac{t}{T_{\mathrm{emp}}} \right)^{m'} \left( 1 - \frac{t}{T_{\mathrm{emp}}} \right)^{n'} \mathrm{d}\left( \frac{t}{T_{\mathrm{emp}}} \right), \tag{43}$$

where $A'$ is the coefficient $A$ for the modified model and is uniquely determined from (43) once $(m', n')$ is given. We examine the two cases $(m', n') = (1, 1)$ and $(0.5, 0.5)$. Here, smaller values of $m', n'$ imply shorter durations during which the unit-time fish count is close to 0. Under the computational conditions in **Section 3.1**, for the nominal case and more fluctuating case $b^{(w)} \to \sqrt{2} b^{(w)}$, as reported in **Table B5**, the sensitivity of Ave and Std against the WT dynamics is clearly higher than that of the nominal case. We also investigate the influences of $\omega$, confirming the higher sensitivity of the unit-time fish count against the biological clock, although not presented here. **Table B6** compares the results for different values of $\kappa_1$. Finally, the eDNA concentration data at Kisuki is provided in **Table B7**.

**Table B1.** Estimated parameter values of the quadruple $(\kappa_0, \kappa_1, \eta, \lambda)$. Yearly average values are also presented. $R^2$ values for the linear regression $\kappa_0 + \kappa_1 t$ are also presented

| Year | $\kappa_0$ (°C) | $\kappa_1$ (°C/day) | $\eta$ (1/day) | $\lambda$ (°C/day$^{1/2}$) | $R^2$ |
|---|---|---|---|---|---|
| 2025 | 6.196E+00 | 1.121E-01 | 2.440E-01 | 9.179E-01 | 0.923 |
| 2024 | 6.988E+00 | 1.106E-01 | 1.276E-01 | 8.801E-01 | 0.879 |
| 2023 | 7.675E+00 | 9.992E-02 | 2.222E-01 | 8.812E-01 | 0.901 |
| 2022 | 6.072E+00 | 1.261E-01 | 1.721E-01 | 9.004E-01 | 0.927 |
| 2021 | 7.532E+00 | 1.017E-01 | 2.422E-01 | 9.051E-01 | 0.913 |
| 2020 | 7.600E+00 | 1.071E-01 | 1.230E-01 | 7.644E-01 | 0.895 |
| 2019 | 7.396E+00 | 1.077E-01 | 1.645E-01 | 7.865E-01 | 0.923 |
| 2018 | 5.882E+00 | 1.195E-01 | 2.036E-01 | 9.122E-01 | 0.930 |
| 2017 | 5.497E+00 | 1.262E-01 | 1.917E-01 | 7.709E-01 | 0.946 |
| 2016 | 6.607E+00 | 1.135E-01 | 1.937E-01 | 8.145E-01 | 0.932 |
| Average | 6.744E+00 | 1.124E-01 | 1.884E-01 | 8.533E-01 | |

**Table B2.** Average (Ave) and standard deviation (Std) of the start date, end date, migration duration, and total fish count for the nominal case, case $a^{(w)} \to +\infty$, case $b^{(w)} \to \sqrt{1/2}b^{(w)}$, and case $b^{(w)} \to \sqrt{2}b^{(w)}$.

| | Nominal | $a^{(w)} \to +\infty$ | $b^{(w)} \to \sqrt{1/2}b^{(w)}$ | $b^{(w)} \to \sqrt{2}b^{(w)}$ |
|---|---|---|---|---|
| Ave of start date (day) | 1.346E+01 | 2.070E+01 | 1.756E+01 | 9.695E+00 |
| Std of start date (day) | 7.725E+00 | 0 | 6.945E+00 | 7.536E+00 |
| Ave of end date (day) | 1.384E+02 | 1.485E+02 | 1.434E+02 | 1.327E+02 |
| Std of end date (day) | 6.321E+00 | 0 | 5.745E+00 | 6.764E+00 |
| Ave of migration duration (day) | 1.250E+02 | 1.278E+02 | 1.258E+02 | 1.230E+02 |
| Std of migration duration (day) | 3.969E+00 | 0 | 2.918E+00 | 5.624E+00 |
| Ave of total fish count (ind) | 8.369E+05 | 8.381E+05 | 8.374E+05 | 8.342E+05 |
| Std of total fish count (ind) | 2.037E+05 | 2.038E+05 | 2.037E+05 | 2.044E+05 |

**Table B3.** Similar to **Table B2** but for different values of $\omega$.

| | $\omega = 1$ | $\omega = 2$ (Nominal) | $\omega = 10$ | $\omega = 100$ |
|---|---|---|---|---|
| Ave of start date (day) | 1.345E+01 | 1.346E+01 | 1.346E+01 | 1.345E+01 |
| Std of start date (day) | 7.715E+00 | 7.725E+00 | 7.720E+00 | 7.718E+00 |
| Ave of end date (day) | 1.413E+02 | 1.384E+02 | 1.362E+02 | 1.357E+02 |
| Std of end date (day) | 7.715E+00 | 6.321E+00 | 6.849E+00 | 7.159E+00 |
| Ave of migration duration (day) | 1.278E+02 | 1.250E+02 | 1.227E+02 | 1.222E+02 |
| Std of migration duration (day) | 2.060E-09 | 3.969E+00 | 7.135E+00 | 7.861E+00 |
| Ave of total fish count (ind) | 8.377E+05 | 8.369E+05 | 8.364E+05 | 8.364E+05 |
| Std of total fish count (ind) | 2.037E+05 | 2.037E+05 | 2.038E+05 | 2.042E+05 |

**Table B4.** Similar to **Table B2** but for the nomina case, the cases $T_{emp} \to (1 \pm 0.1) T_{emp}$, and randomized $T_{emp}$.

| | Nominal | $T_{emp} \to 1.1 T_{emp}$ | $T_{emp} \to 0.9 T_{emp}$ | Random $T_{emp}$ |
|---|---|---|---|---|
| Ave of start date (day) | 1.346E+01 | 1.347E+01 | 1.346E+01 | 1.347E+01 |
| Std of start date (day) | 7.725E+00 | 7.724E+00 | 7.726E+00 | 7.723E+00 |
| Ave of end date (day) | 1.384E+02 | 1.466E+02 | 1.278E+02 | 1.379E+02 |
| Std of end date (day) | 6.321E+00 | 6.064E+00 | 7.141E+00 | 9.166E+00 |
| Ave of migration duration (day) | 1.250E+02 | 1.332E+02 | 1.144E+02 | 1.244E+02 |
| Std of migration duration (day) | 3.969E+00 | 5.628E+00 | 1.888E+00 | 7.676E+00 |
| Ave of total fish count (ind) | 8.369E+05 | 9.177E+05 | 7.538E+05 | 8.361E+05 |
| Std of total fish count (ind) | 2.037E+05 | 2.242E+05 | 1.834E+05 | 2.120E+05 |

**Table B5.** Similar to **Table B2** but for different values of $(m,n)$.

| | $m=n=1$ | $m=n=1$ ($b^{(w)} \to \sqrt{2}b^{(w)}$) | $m=n=0.5$ | $m=n=0.5$ ($b^{(w)} \to \sqrt{2}b^{(w)}$) |
|---|---|---|---|---|
| Ave of start date (day) | 1.346E+01 | 9.695E+00 | 1.346E+01 | 9.695E+00 |
| Std of start date (day) | 7.725E+00 | 7.536E+00 | 7.725E+00 | 7.536E+00 |
| Ave of end date (day) | 1.384E+02 | 1.327E+02 | 1.384E+02 | 1.327E+02 |
| Std of end date (day) | 6.321E+00 | 6.764E+00 | 6.321E+00 | 6.764E+00 |
| Ave of migration duration (day) | 1.250E+02 | 1.230E+02 | 1.250E+02 | 1.230E+02 |
| Std of migration duration (day) | 3.969E+00 | 5.624E+00 | 3.969E+00 | 5.624E+00 |
| Ave of total fish count (ind) | 8.295E+05 | 8.219E+05 | 8.255E+05 | 8.157E+05 |
| Std of total fish count (ind) | 2.037E+05 | 2.046E+05 | 2.041E+05 | 2.052E+05 |

**Table B6.** Similar to **Table B2** but for the nominal case and cases $\hat{\kappa} \to (1 \pm 0.1)\hat{\kappa}$.

| | Nominal | $\hat{\kappa} \to 1.1\hat{\kappa}$ | $\hat{\kappa} \to 0.9\hat{\kappa}$ |
|---|---|---|---|
| Ave of start date (day) | 1.346E+01 | 1.297E+01 | 1.400E+01 |
| Std of start date (day) | 7.725E+00 | 7.315E+00 | 8.178E+00 |
| Ave of end date (day) | 1.384E+02 | 1.341E+02 | 1.411E+02 |
| Std of end date (day) | 6.321E+00 | 5.666E+00 | 7.560E+00 |
| Ave of migration duration (day) | 1.250E+02 | 1.212E+02 | 1.271E+02 |
| Std of migration duration (day) | 3.969E+00 | 5.194E+00 | 2.058E+00 |
| Ave of total fish count (ind) | 8.369E+05 | 8.344E+05 | 8.375E+05 |
| Std of total fish count (ind) | 2.037E+05 | 2.040E+05 | 2.038E+05 |

**Table B7.** The eDNA concentration data at Kisuki in 2025.

| Date | Concentration (copies/ml) |
|---|---|
| April 3 | 0* |
| April 9 | 0* |
| April 17 | 7.73 |
| April 24 | 4.71 |
| April 30 | 13.31 |
| May 8 | 18.39 |
| May 14 | 48.65 |
| May 21 | 165.49 |
| May 28 | 135.11 |
| June 5 | 171.22 |
| June 12 | 271.13 |
| June 19 | 117.45 |
| July 2 | 56.83 |
| July 9 | 48.05 |
| July 17 | 76.78 |
| July 24 | 48.27 |
| July 31 | 30.91 |
| August 6 | 47.85 |
| August 15 | 27.06 |
| August 21 | 72.96 |
| August 28 | 102.20 |
| September 4 | 112.58 |
| September 12 | 168.92 |
| September 24 | 305.68 |
| October 1 | 165.98 |
| October 9 | 113.49 |
| October 14 | 34.36 |
| October 23 | 19.61 |
| October 29 | 9.88 |
| November 6 | 1.98 |
| November 12 | 1.33 |

* Not detected.

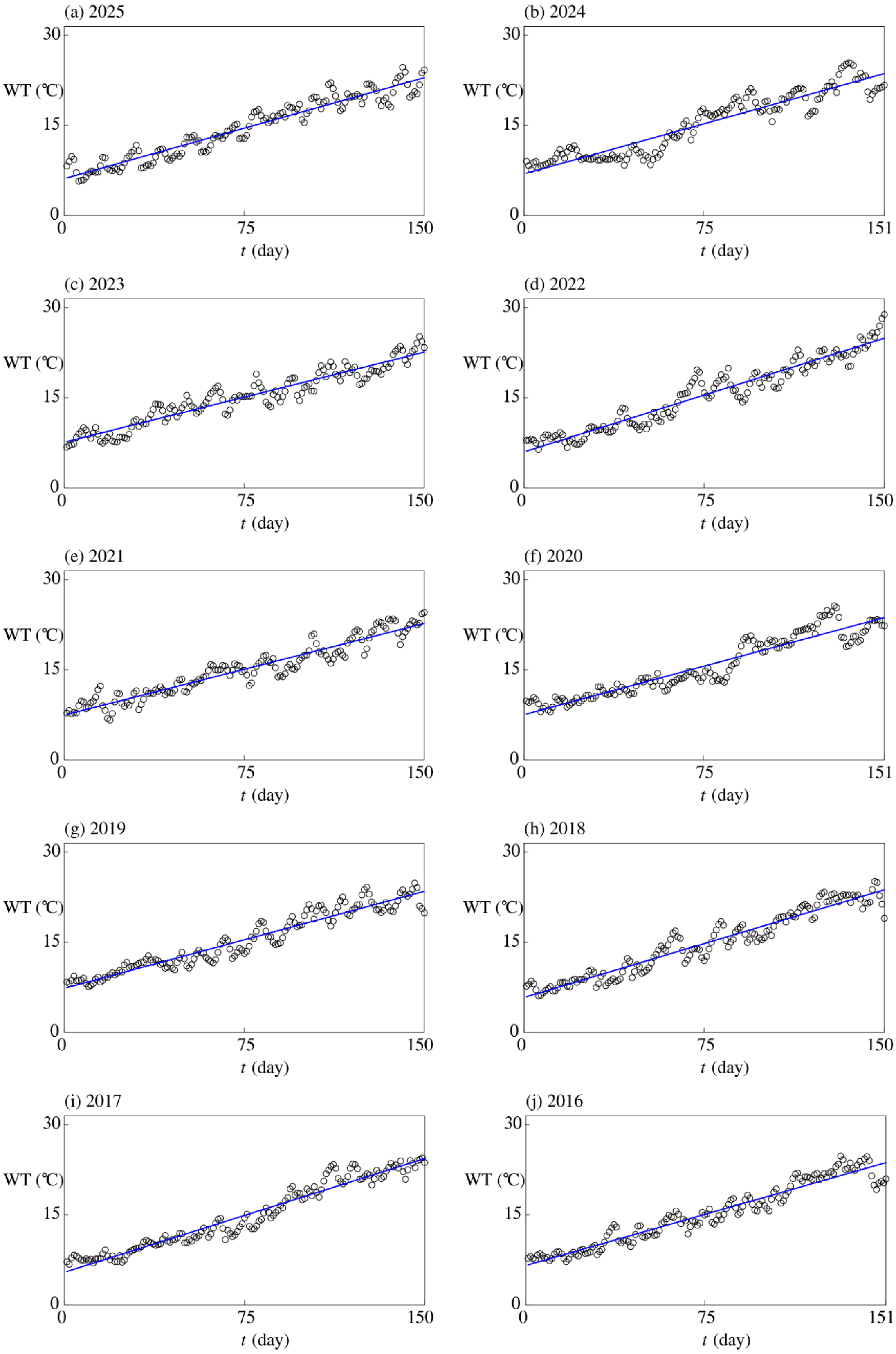

**Figure B1.** Visualizations of the linear regression of daily WT (°C) each $t$ (day) from February 1 (day 1) to June 30 (day 150 or 151): empirical values (black) and regression results (blue).

**Reference**

[1] Bernhardt, J. R., & O'Connor, M. I. (2021). Aquatic biodiversity enhances multiple nutritional benefits to humans. Proceedings of the National Academy of Sciences, 118(15), e1917487118. https://doi.org/10.1073/pnas.1917487118

[2] Lucas, M. C., Baras, E., Thom, T. J., Duncan, A., & Slavík, O. (2001). Migration of freshwater fishes, Blackwell Science, Oxford.

[3] Cooke, S. J., Bergman, J. N., Twardek, W. M., Piczak, M. L., Casselberry, G. A., Lutek, K., ... & Lennox, R. J. (2022). The movement ecology of fishes. Journal of Fish Biology, 101(4), 756-779. https://doi.org/10.1111/jfb.15153

[4] Kurasawa, A., Onishi, Y., Koba, K., Fukushima, K., & Uno, H. (2024). Sequential migrations of diverse fish community provide seasonally prolonged and stable nutrient inputs to a river. Science advances, 10(43), eadq0945. https://doi.org/10.1126/sciadv.adq0945

[5] Towne, K., Olinger, C. T., & Wright, G. (2026). Comparison of the electrified dozer trawl and boat electrofishing for fish community sampling in turbid and clear river systems. North American Journal of Fisheries Management, vqag029. https://doi.org/10.1093/najfmt/vqag029

[6] Fujihara, M., Watanabe, K., Yoshioka, H., Ichion, E., Chono, S., & Izumi, T. (2026). Computational model for upstream migration of Ayu (*Plecoglossus altivelis*) in an agricultural canal equipped with hydraulic drops and tilting weirs. Paddy and Water Environment, 24(1), 1-11. https://doi.org/10.1007/s10333-025-01046-3

[7] Ramírez-Álvarez, R., Contreras, S., Vivancos, A., Reid, M., López-Rodríguez, R., & Górski, K. (2022). Unpacking the complexity of longitudinal movement and recruitment patterns of facultative amphidromous fish. Scientific Reports, 12(1), 3164. https://doi.org/10.1038/s41598-022-06936-8

[8] Quade, A. H., Ferrara, A., Fontenot, Q., Boyle, K. S., David, S. R., & Rieucau, G. (2025). Spotting gar using imaging sonar: The effects of river–floodplain habitat connectivity on a lepisosteid assemblage. Transactions of the American Fisheries Society, 154(2), 115-126. https://doi.org/10.1093/tafafs/vnae006

[9] Abe, K., Kitanishi, R., Habe, H., Otani, M., & Iguchi, N. (2025). A Fish Counting System with Video Camera for Hatchery-produced Juvenile Fish. IEICE Transactions on Information and Systems, 2024EDP7282. https://doi.org/10.1587/transinf.2024EDP7282

[10] Rourke, M. L., Fowler, A. M., Hughes, J. M., Broadhurst, M. K., DiBattista, J. D., Fielder, S., ... & Furlan, E. M. (2022). Environmental DNA (eDNA) as a tool for assessing fish biomass: A review of approaches and future considerations for resource surveys. Environmental DNA, 4(1), 9-33. https://doi.org/10.1002/edn3.185

[11] Kleiman, L. E., Young, J. M., Rhoades, O. K., Ajemian, M. J., & Baldwin, J. D. (2026). The role of freshwater discharges on Common Snook movements and distribution in a south Florida estuary. Marine and Coastal Fisheries, 18(2), vtag002. https://doi.org/10.1093/mcfafs/vtag002

[12] Rigby, C. R., & Gaeta, J. W. (2026). Evaluating Patterns of Juvenile Chinook Salmon Out-Migration During Pulse Flow Events in California's Sacramento River to Inform Operations of a Newly Proposed Water Infrastructure Project. River Research and Applications. Online published. https://doi.org/10.1002/rra.70155Digital

[13] Al-Chokhachy, R., Peka, R., Horgen, E., Kaus, D. J., Loux, T., & Heki, L. (2022). Water availability drives instream conditions and life-history of an imperiled desert fish: a case study to inform water management. Science of The Total Environment, 832, 154614. https://doi.org/10.1016/j.scitotenv.2022.154614

[14] Paíz, R., Thomas, R. Q., Carey, C. C., de Eyto, E., Jones, I. D., Delany, A. D., ... & Jennings, E. (2025). Near-term lake water temperature forecasts can be used to anticipate the ecological dynamics of freshwater species. Ecosphere, 16(7), e70335. https://doi.org/10.1002/ecs2.70335

[15] Potapova, A., Moore, J. W., Parken, C. K., Sloat, M., Atlas, W., & Jones, L. A. (2026). Range-wide life history diversity and climate exposure in adult Chinook salmon. Ecosphere, 17(3), e70572. https://doi.org/10.1002/ecs2.70572

[16] Weedop, D., Romer, J. D., Ziller, J. S., & Murphy, C. A. (2026). Timing is everything: Drivers of upstream movement of fishes. Transactions of the American Fisheries Society, vnag008. https://doi.org/10.1093/tafafs/vnag008

[17] Keefer, M. L., & Caudill, C. C. (2026). Migration phenology of adult Chinook salmon: tradeoffs among acute and cumulative thermal exposure risks. Journal of Thermal Biology, 104388. https://doi.org/10.1016/j.jtherbio.2026.104388

[18] Moore, J. W., Ulaski, M. E., Wilson, K. L., Martin, T. G., Kuiper, S. D., Peacock, S. J., ... & Zdasiuk, B. J. (2025). A safe operating space for Salmon watersheds under rapid climate change. Fish and Fisheries, 26(6), 1213-1228. https://doi.org/10.1111/faf.70027

[19] Meixler, M. S., Bain, M. B., & Walter, M. T. (2009). Predicting barrier passage and habitat suitability for migratory fish species. Ecological Modelling, 220(20), 2782-2791. https://doi.org/10.1016/j.ecolmodel.2009.07.014

[20] Vervaat, W. (1979). A relation between Brownian bridge and Brownian excursion. The Annals of Probability, 143-149. https://doi.org/10.1214/aop/1176995155

[21] Kreuser, A. M., Pendleton, D. E., Record, N. R., & Meyer-Gutbrod, E. L. (2026). Individual movement modeling expands the power of migratory species observations: North Atlantic right whale case study. Limnology and Oceanography: Methods, e70049. https://doi.org/10.1002/lom3.70049

[22] Jorzik, I., Fiedler, W., Kaatz, M., Rotics, S., Nathan, R., Wikelski, M., & Flack, A. (2026). Timing shapes flyway selection in juvenile white storks at the European migratory divide. Journal of Ornithology, 167(1), 131-142. https://doi.org/10.1007/s10336-025-02322-z

[23] Acolas, M. L., Le Pichon, C., & Rochard, E. (2017). Spring habitat use by stocked one year old European sturgeon Acipenser sturio in the freshwater-oligohaline area of the Gironde estuary. Estuarine, Coastal and Shelf Science, 196, 58-69. https://doi.org/10.1016/j.ecss.2017.06.029

[24] Swadling, D. S., Knott, N. A., Taylor, M. D., Rees, M. J., Cadiou, G., & Davis, A. R. (2024). Consequences of juvenile fish movement and seascape connectivity: does the concept of nursery habitat need a rethink?. Estuaries and Coasts, 47(3), 607-621. https://doi.org/10.1007/s12237-023-01323-6

[25] Futia, M. H., Binder, T. R., Henderson, M., & Marsden, J. E. (2024). Modeling regional occupancy of fishes using acoustic telemetry: A model comparison framework applied to lake trout. Animal Biotelemetry, 12(1), 25. https://doi.org/10.1186/s40317-024-00380-3

[26] Abi Jaber, E., & Villeneuve, S. (2025). Gaussian agency problems with memory and linear contracts. Finance and Stochastics, 29(1), 143-176. https://doi.org/10.1007/s00780-024-00548-y

[27] Singham, D. I., & Rodgers, M. (2022). A shortage probability metric for battery depletion risk. Operations Research Letters, 50(6), 660-666. https://doi.org/10.1016/j.orl.2022.10.005

[28] Bischoff, T., & Deck, K. (2024). Unpaired downscaling of fluid flows with diffusion bridges. Artificial Intelligence for the Earth Systems, 3(2), e230039. https://doi.org/10.1175/AIES-D-23-0039.1

[29] Qian, Z., Süli, E., & Zhang, Y. (2022). Random vortex dynamics via functional stochastic differential equations. Proceedings of the Royal Society A: Mathematical, Physical and Engineering Sciences, 478(2266). https://doi.org/10.1098/rspa.2022.0030

[30] Lenner, N., Häring, M., Eule, S., Großhans, J., & Wolf, F. (2026). Landscape of Target State Directed Processes in Living Systems. PRX Life, 4(1), 013024. https://doi.org/10.1103/cx3h-q6d2

[31] Amankwah, M., Bersani, A., Calvetti, D., Davico, G., Somersalo, E., & Viceconti, M. (2024). Exploring muscle recruitment by Bayesian methods during motion. Chaos, Solitons & Fractals, 185, 115082. https://doi.org/10.1016/j.chaos.2024.115082

[32] Alazemi, F., Alsenafi, A., Chen, Y., & Zhou, H. (2024). Parameter estimation for the complex fractional Ornstein–Uhlenbeck processes with Hurst parameter H∈(0, 12). Chaos, Solitons & Fractals, 188, 115556. https://doi.org/10.1016/j.chaos.2024.115556

[33] Behjoo, H., & Chertkov, M. M. (2024). Space-Time Diffusion Bridge. IFAC-PapersOnLine, 58(17), 274-279. https://doi.org/10.1016/j.ifacol.2024.10.181

[34] Nobis, G., Springenberg, M., Belova, A., Daems, R., Knochenhauer, C., Opper, M., ... & Samek, W. (2026). Fractional diffusion bridge models. Advances in Neural Information Processing Systems, 38, 114521-114571. https://proceedings.neurips.cc/paper_files/paper/2025/file/a66caa1703fe34705a4368c3014c1966-Paper-Conference.pdf

[35] Yoshioka, H. (2025). CIR bridge for modeling of fish migration on sub-hourly scale. Chaos, Solitons & Fractals, 199, 116874. https://doi.org/10.1016/j.chaos.2025.116874

[36] Louriki, M. (2022). Brownian bridge with random length and pinning point for modelling of financial information. Stochastics, 94(7), 973-1002. https://doi.org/10.1080/17442508.2021.2017438

[37] Yoshioka, H. (2026a). A minimization principle behind the diffusion bridge of diurnal fish migration, Mathematical Biosciences, 396, 109684. https://doi.org/10.1016/j.mbs.2026.109684

[38] Yoshioka, H. (2026b). Multiple timescales in collective motion: daily and intraday upstream fish migration focusing on Feller condition. Physica A. In press. http://arxiv.org/abs/2602.06606

[39] Monthus, C., & Mazzolo, A. (2022). Conditioned diffusion processes with an absorbing boundary condition for finite or infinite horizon. Physical Review E, 106(4), 044117. DOI: https://doi.org/10.1103/PhysRevE.106.044117

[40] Baker, E. L., Yang, G., Severinsen, M. L., Hipsley, C. A., & Sommer, S. (2024). Conditioning non-linear and infinite-dimensional diffusion processes. Advances in Neural Information Processing Systems, 37, 10801-10826. https://proceedings.neurips.cc/paper_files/paper/2024/file/14ad9256c430e6c8977e470d8e268320-Paper-Conference.pdf

[41] Zhang, R., Huang, Y., Cao, Y., & Wang, H. (2026). Molebridge: Synthetic space projecting with discrete markov bridges. Advances in Neural Information Processing Systems, 38, 162043-162062. https://proceedings.neurips.cc/paper_files/paper/2025/file/ed2e1bfa8721d7cb04b8f6f92ab8c0ba-Paper-Conference.pdf

[42] Chen, Y., & Georgiou, T. (2015). Stochastic bridges of linear systems. IEEE Transactions on Automatic Control, 61(2), 526-531. https://doi.org/10.1109/TAC.2015.2440567

[43] Mazzolo, A. (2017). Constraint Ornstein-Uhlenbeck bridges. Journal of Mathematical Physics, 58(9). https://doi.org/10.1063/1.5000077

[44] Watanabe, S., Iida, M., Lord, C., Keith, P., & Tsukamoto, K. (2014). Tropical and temperate freshwater amphidromy: a comparison between life history characteristics of Sicydiinae, ayu, sculpins and galaxiids. Reviews in Fish Biology and Fisheries, 24(1), 1-14. https://doi.org/10.1007/s11160-013-9316-8

[45] Sakai, A. (2010). Forecast of the first ascending day and number of ascending ayu *Plecoglossus altivelis altivelis* in rivers flowing into Lake Biwa. Nippon Suisan Gakkaishi, 76(4), 670-677. In Japanese with English Abstract. https://doi.org/10.2331/suisan.76.670

[46] Tago, Y. (2002). Migration behaviors of sea-run ayu *Plecoglossus altivelis* (Pisces) in Toyama Bay, Japan. Nippon Suisan Gakkaishi, 68(4), 554-563. In Japanese with English Abstract. https://doi.org/10.2331/suisan.68.554

[47] Yates, M. C., Wilcox, T. M., Kay, S., Peres-Neto, P., & Heath, D. D. (2025). A framework to unify the relationship between numerical abundance, biomass, and environmental DNA. Environmental DNA, 7(2), e70073. https://doi.org/10.1002/edn3.70073

[48] Karatzas, I., & Shreve, S. (2014). Brownian motion and stochastic calculus. Springer, New York.

[49] Hefter, M., & Herzwurm, A. (2018). Strong convergence rates for Cox–Ingersoll–Ross processes—full parameter range. Journal of Mathematical Analysis and Applications, 459(2), 1079-1101. https://doi.org/10.1016/j.jmaa.2017.10.076

[50] Abi Jaber, E. (2025). Simulation of square-root processes made simple: applications to the Heston model. arXiv preprint arXiv:2412.11264.

[51] Abi Jaber, E., & Attal, E. (2025). Simulating integrated Volterra square-root processes and Volterra Heston models via Inverse Gaussian. arXiv preprint arXiv:2504.19885.

[52] Abi Jaber, E., Attal, E., & Rosenbaum, M. (2025). From Hyper Roughness to Jumps as $ H¥to-1/2$. arXiv preprint arXiv:2503.16985.

[53] Tsukamoto, K., & Uchida, K. (1992). Migration Mechanism of the ayu, in Oceanic and Anthropogenic Controls of Life in the Pacific Ocean, Ilyichev, V.I. and Anikiev, V.V., eds., Springer, Dordrecht, 145-172. 10.1007/978-94-011-2773-8_12

[54] Tajima, A. (2023). Historical overview of poultry in Japan. The Journal of Poultry Science, 60(2), 2023015. https://doi.org/10.2141/jpsa.2023015

[55] Kuroda, K., Mori, T., Yuasa, H., & Uchida, K. (2026). Recovery of plant species and functional diversity from anthropogenic impacts on the river embankment. Plant Ecology, 227(3), 29. https://doi.org/10.1007/s11258-025-01590-2

[56] Mousumi, K. A., Onishi, T., Sueyoshi, M., Harada, M., & Hiramatsu, K. (2025). River water temperature variability: classification, diurnal change, and precipitation impacts in Nagara River tributaries. Hydrological Research Letters, 19(4), 327-334. https://doi.org/10.3178/hrl.25-00031

[57] Nagayama, S., Ohta, T., Fujii, R., Harada, M., & Iizuka, T. (2025). Habitat use and growth strategies of amphidromous fish "ayu" throughout a river system. Scientific Reports, 15(1), 18695. https://doi.org/10.1038/s41598-025-02988-8

[58] Aino, S., Yodo, T., & Yoshioka, M. (2015). Changes in the composition of stock origin and standard length of ayu *Plecoglossus altivelis altivelis* during the Tomozuri angling season in the Nagara River, central Japan. Fisheries science, 81(1), 37-42. https://doi.org/10.1007/s12562-014-0822-y

[59] Iritani, R., Oshima, K., Oda, T., Sasaki, O., Oki, K., Miyamoto, H., & Hirabayashi, Y. (2025). Historical water temperature change in Japan over the past several decades. Hydrological Research Letters, 19(4), 260-267. https://doi.org/10.3178/hrl.25-00018

[60] Itsukushima, R., Ohtsuki, K., & Sato, T. (2024). Drivers of rising monthly water temperature in river estuaries. Limnology and Oceanography, 69(3), 589-603. https://doi.org/10.1002/lno.12507

[61] Yamanaka, H., & Minamoto, T. (2016). The use of environmental DNA of fishes as an efficient method of determining habitat connectivity. Ecological Indicators, 62, 147-153. https://doi.org/10.1016/j.ecolind.2015.11.022

[62] Kawaida, S., Horinouchi, M., Kurata, K., & Toda, K. (2025). Differences in benthic assemblage structures between vegetated and unvegetated habitats in a salt marsh in Lake Shinji, western Japan. Fisheries Science, 91(1), 33-46. https://doi.org/10.1007/s12562-024-01831-9

[63] Yajima, H., & Yoshioka, Y. (2025). Response of interconnected estuarine lakes to sea-level rise and large hydrological structures. Journal of Hydrology: Regional Studies, 59, 102432. https://doi.org/10.1016/j.ejrh.2025.102432

[64] Asaeda, T., García de Jalón, D., O'Hare, M., & Rahman, M. (2025). Sequential riparian vegetation alteration in Japanese river landscapes. International Journal of River Basin Management, 23(1), 183-195. https://doi.org/10.1080/15715124.2023.2273843

[65] Kotani, T., Gotoh, T., & Fukuoka, S. (2024). Development of an integrated analysis method for rainfall runoff, flood flow and sediment transport in the Hii river and its application for basin-wide flood control. Advances in River Engineering, 30, 453-458. In Japanese. https://doi.org/10.11532/river.30.0_453

[66] Yoshioka, H., Tsujimura, M., & Yoshioka, Y. (2026). Fishery resources management. Chapter 6 in the book: Yoshioka H., Tsujimura M. (Eds.): Stochastic Analysis and Control in Energy, Environmental and Resource Management Applications, Elsevier.

[67] Morgado-Gamero, W. B., Tournayre, O., & Cristescu, M. E. (2025). Comparative Decay Dynamics and Detectability of eDNA and eRNA in Connected and Isolated Freshwater Mesocosms Using Digital PCR. Molecular Ecology Resources, 25(8), e70028. https://doi.org/10.1111/1755-0998.70028

[68] Snyder, E. D., Tank, J. L., Pruitt, A. N., Peters, B., Brandão-Dias, P. F., Curtis, E. M., ... & Lamberti, G. A. (2025). Warming Increases Environmental DNA (eDNA) Removal Rates in Flowing Waters. Environmental DNA, 7(3), e70094. https://doi.org/10.1002/edn3.70094
[69] Doi, H., Inui, R., Akamatsu, Y., Kanno, K., Yamanaka, H., Takahara, T., & Minamoto, T. (2017). Environmental DNA analysis for estimating the abundance and biomass of stream fish. Freshwater Biology, 62(1), 30-39. https://doi.org/10.1111/fwb.12846
[70] Talarico, L., Petrosino, G., Rossi, A. R., Franchini, P., & Tancioni, L. (2026). Controlled Experiments Reveal Moderate, Nonlinear Relationships Between eDNA Concentration and Fish Biomass in Three Freshwater Species of Monitoring Relevance. Ecology and Evolution, 16(2), e73129. https://doi.org/10.1002/ece3.73129
[71] Henze, N. (2024). Asymptotic Stochastics. Springer, Berlin, Heidelberg.
[72] Jo, T. S., Murakami, H., & Nakadai, R. (2025). Spatial dispersal of environmental DNA particles in lentic and marine ecosystems: An overview and synthesis. Ecological Indicators, 174, 113469. https://doi.org/10.1016/j.ecolind.2025.113469
[73] Perry, W. B., Milner, N., Jâms, I. B., de Bruyn, M., Ormerod, S. J., Deiner, K., ... & Creer, S. (2025). Quantitative insights into the spatio-temporal variation of Atlantic salmon (Salmo salar) biomass in a river catchment using eDNA metabarcoding. Journal of Fish Biology. Published online. https://doi.org/10.1111/jfb.70172
[74] Jo, T. S., & Doi, H. (2026). Does Allometric Scaling Improve Estimates of Population Abundance Based on Environmental DNA?. Molecular Ecology, 35(6), e70303. https://doi.org/10.1111/mec.70303
[75] Yoshioka, H. (2026). A correction note to a part of a proof in: CIR bridge for modeling of fish migration on sub-hourly scale (Chaos, Solitons and Fractals 199 (2025) 116874). Chaos, Solitons & Fractals. To appear.